\documentclass[prb,twocolumn,showpacs,preprintnumbers,amsmath,amssymb,superscriptaddress,floatfix]{revtex4}

\usepackage{graphicx}
\usepackage[T1]{fontenc}
\usepackage{lmodern}
\usepackage[latin1]{inputenc}
\usepackage{dcolumn}
\usepackage{bm}
\usepackage{color}
\usepackage[colorlinks,plainpages=false,linkcolor=blue,urlcolor=blue,citecolor=black,pdfpagemode=UseNone,pdfstartview=FitBH]{hyperref}
\makeatletter\renewcommand{\fnum@figure}[1]{\figurename~\thefigure~(color online).}\makeatother

\begin{document}

\title{Orbital superexchange and crystal field simultaneously at play in $\mbox{YVO}_{3}$:
\\
resonant inelastic x-ray scattering at the V $L$ edge and the O $K$ edge}

\author{E.~Benckiser}
\email[]{e.benckiser@fkf.mpg.de}
\affiliation{II.~Physikalisches Institut, Z\"{u}lpicher Str.\ 77, Universit\"{a}t zu K\"{o}ln, 50937 K\"{o}ln, Germany}
\affiliation{Max Planck Institute for Solid State Research, Heisenbergstra{\ss}e 1, 70569 Stuttgart, Germany}

\author{L.~Fels}
\affiliation{II.~Physikalisches Institut, Z\"{u}lpicher Str.\ 77, Universit\"{a}t zu K\"{o}ln, 50937 K\"{o}ln, Germany}

\author{G.~Ghiringhelli}
\affiliation{CNR/SPIN - Dipartimento di Fisica, Politecnico di Milano, piazza Leonardo da Vinci 32, 20133 Milano, Italy}

\author{M.~Moretti Sala}
\affiliation{CNR/SPIN - Dipartimento di Fisica, Politecnico di Milano, piazza Leonardo da Vinci 32, 20133 Milano, Italy}
\affiliation{European Synchrotron Radiation Facility, BP 220, 38043 Grenoble cedex, France}

\author{T.~Schmitt}
\affiliation{Swiss Light Source, Paul Scherrer Institut, CH-5232 Villigen PSI, Switzerland}

\author{J.~Schlappa}
\affiliation{Swiss Light Source, Paul Scherrer Institut, CH-5232 Villigen PSI, Switzerland}
\affiliation{Institut Methoden und Instrumentierung der Synchrotronstrahlung, BESSY II, Albert-Einstein-Str. 15,
12489 Berlin, Germany}

\author{V.~N.~Strocov}
\affiliation{Swiss Light Source, Paul Scherrer Institut, CH-5232 Villigen PSI, Switzerland}

\author{N.~Mufti}
\affiliation{Department of Chemical Physics, Zernike Institute for Advanced Materials, University of Groningen,
Nijenborgh 4, 9747 AG Groningen, The Netherlands}
\affiliation{Department of Physics, State University of Malang, Jl.\ Semarang No. 5, 65145 Malang, Indonesia}

\author{G.~R.~Blake}
\affiliation{Department of Chemical Physics, Zernike Institute for Advanced Materials, University of Groningen,
Nijenborgh 4, 9747 AG Groningen, The Netherlands}

\author{A.~A.~Nugroho}
\affiliation{Department of Chemical Physics, Zernike Institute for Advanced Materials, University of Groningen,
Nijenborgh 4, 9747 AG Groningen, The Netherlands}
\affiliation{Jurusan Fisika, Institut Teknologi Bandung, Jl. Ganesha 10, Bandung 40132, Indonesia}

\author{T.~T.~M.~Palstra}
\affiliation{Department of Chemical Physics, Zernike Institute for Advanced Materials, University of Groningen,
Nijenborgh 4, 9747 AG Groningen, The Netherlands}

\author{M.~W.~Haverkort}
\affiliation{Max Planck Institute for Solid State Research, Heisenbergstra{\ss}e 1, 70569 Stuttgart, Germany}
\affiliation{Department of Physics and Astronomy, University of British Columbia, Vancouver, Canada V6T1Z1}

\author{K.~Wohlfeld}
\affiliation{Institute for Theoretical Solid State Physics, IFW Dresden, D-01069 Dresden, Germany}
\affiliation{Stanford Institute for Materials and Energy Sciences, SLAC National Accelerator Laboratory, Menlo Park, California 94025, USA}

\author{M.~Gr\"{u}ninger}
\email[]{grueninger@ph2.uni-koeln.de}
\affiliation{II.~Physikalisches Institut, Z\"{u}lpicher Str.\ 77, Universit\"{a}t zu K\"{o}ln, 50937 K\"{o}ln, Germany}

\date{\today}

\begin{abstract}
We report on the observation of orbital excitations in YVO$_3$ by means of resonant inelastic x-ray scattering (RIXS) at energies across the vanadium \textit{L}$_{3}$ and oxygen \textit{K} absorption edges. Due to the excellent experimental resolution we are able to resolve the intra-$t_{2g}$ excitations at 0.1\,-\,0.2\,eV, 1.07\,eV, and 1.28\,eV, the lowest excitations from the $t_{2g}$ into the $e_g$ levels at 1.86\,eV, and further excitations above 2.2\,eV.\@ For the intra-$t_{2g}$ excitations at 0.1\,-\,0.2\,eV, the RIXS peaks show small shifts of the order of 10\,-\,40\,meV as a function of temperature and of about 13\,-\,20\,meV as a function of the transferred momentum $q \! \parallel \! a$. We argue that the latter reflects a finite dispersion of the orbital excitations. For incident energies tuned to the oxygen \textit{K} edge, RIXS is more sensitive to intersite excitations. We observe excitations across the Mott-Hubbard gap and find an additional feature at 0.4\,eV which we attribute to two-orbiton scattering, i.e., an exchange of orbitals between adjacent sites. Altogether, these results indicate that both superexchange interactions and the coupling to the lattice are important for a quantitative understanding of the orbital excitations in YVO$_3$.
\end{abstract}


\maketitle

\section{Introduction}

The electronic properties of strongly correlated tran\-sition-metal oxides strongly depend on spin and orbital degrees of freedom.\cite{Tokura2000} Both spins and orbitals on neighboring sites interact with each other via superexchange interactions,\cite{Kugel1973} giving rise to a complex interplay and potentially to novel states induced by orbital quantum fluctuations. In this context the occurrence of orbital order and more exotic states, like an orbital Peierls state\cite{Ulrich2003,Sirker2003,Horsch2003} and an orbital-liquid ground state have been discussed.\cite{Khaliullin2005} One possibility to test materials for those orbital states is to investigate the character of the elementary orbital excitations. If the superexchange interaction dominates, one expects novel collective elementary excitations, namely orbital waves or orbitons with a significant dispersion,\cite{Ishihara2000} analogous to spin waves in a magnetically ordered state. However, the orbitals are also strongly coupled to the lattice,\cite{Jahn1937} therefore orbital excitations in many compounds are well described in the limit of \lq{}local\rq{} crystal-field excitations. If both, superexchange interactions and the coupling to the lattice are relevant, the spectral signatures become more difficult to interpret, a complex situation that has hardly been studied thus far.\cite{Brink01,Schmidt07,Krivenko12,Horsch2008}

The experimental observation of orbitons at low energies $\leq$\,250\,meV has been claimed based on Raman data of LaMnO$_3$, $R$TiO$_3$, and $R$VO$_3$ with $R$\,=\,rare earth.\cite{Saitoh,Miyasaka2005,Miyasaka2006,Sugai2006,Ulrich2006} However, the orbiton interpretation of these data has caused controversy.\cite{Grueninger2002a,Rueckamp2005,Iliev2007,Benckiser2008,Jandl2010,Miyasaka2005,Miyasaka2006,Sugai2006,Ulrich2006} For manganites with partially occupied $e_g$ states in a predominantly octahedral crystal field, it is meanwhile well established that the orbital degree of freedom is quenched by the strong crystal-field splitting of the order of 1\,eV.\cite{Kovaleva2004,Goessling2008} Recently, it has been shown by resonant inelastic x-ray scattering (RIXS) on 1D cuprate chains that such high-energy orbital excitations may show a significant dispersion.\cite{Wohlfeld11,Schlappa12} The large dispersion reflects the strong superexchange coupling of the cuprates, and the high excitation energy results from the crystal-field splitting. At the same time, the high excitation energy of the lowest orbital excitation is a clear signature that orbital fluctuations are weak in the ground state, and that this orbitally ordered ground state is not affected by the high-energy dispersion.
In contrast, the crystal-field splitting is much weaker for the partially occupied $t_{2g}$ states in titanates and vanadates. For $R$VO$_3$ with $t_{2g}^2$
electron configuration, it has been suggested that orbital quantum fluctuations may be comparatively strong because superexchange interactions between $t_{2g}$ electrons are frustrated on a cubic or nearly cubic lattice.\cite{deRay07,Ishihara2004,Ulrich2003,Horsch2003,Khaliullin2001} From magnetic neutron scattering, indications for highly unusual orbital correlations were found and the existence of an orbital Peierls state in the intermediate phase of YVO$_3$ has been proposed.\cite{Ulrich2003} However, a recent study of the optical excitations across the Mott-Hubbard gap rules out that orbital fluctuations are strong in $R$VO$_3$.\cite{Reul12} Some of us reported the observation of orbital excitations in the optical conductivity of YVO$_3$.\cite{Benckiser2008} An optical absorption band at 0.4\,eV for polarization of the electric field $E$\,$\parallel$\,$c$ has been attributed to the exchange of orbitals on adjacent sites, i.e., to the direct excitation of two orbitons. A two-orbiton interpretation has also been discussed for the RIXS data of the orbital excitations of the $t_{2g}^1$ compounds LaTiO$_3$ and YTiO$_3$.\cite{Ulrich2009,Ament2010}
In the titanates, the two-orbiton character is difficult to prove since the orbital excitations are observed at the same energy of 0.20\,-\,0.25\,eV in RIXS, Raman spectroscopy, and infrared absorption,\cite{Ulrich2006,Ulrich2008,Ulrich2009,Rueckamp2005} while there is no clear signature of a one-orbiton excitation
at about half the energy of the proposed two-orbiton peak. It has been claimed that the RIXS intensity of two-orbiton excitations may exceed that of single-orbiton modes in titanates, in particular if quantum fluctuations are large in the orbital sector.\cite{Ulrich2009,Ament2010}

RIXS is the analogue of Raman spectroscopy with x-ray photons. RIXS is very sensitive to orbital excitations, in particular in the soft x-ray range.\cite{Schmitt2002,Schmitt2004,Schmitt2004b,Ghir04,Chiu05,Ghir06,Duda06,Brai07,Ulrich2008,Hagu08,Forte08a,Ament11,Marra2012} At the V $L$ edge, a photon is resonantly absorbed, exciting a $2p$ core electron to an empty $3d$ state, and subsequently this intermediate state decays by re-emitting a photon, leaving the system in, e.g., an orbitally excited state. One major advantage of RIXS over Raman scattering with visible light is the short wavelength of x-rays, allowing to study the excitations as a function of the transferred momentum $q$. Furthermore, by tuning the energy of the incoming light to a particular absorption band, special ions are selected and potentially also specific crystallographic sites. Since the probing depth for soft x-rays is comparatively large, bulk properties are probed.
Here, we report on RIXS measurements with a resolution of 60\,meV at the V $L_3$ edge and 70\,meV at the O $K$ edge. We observe spin-conserving intra-$t_{2g}$ excitations at 0.1\,-\,0.2\,eV, orbital excitations from the high-spin $S$\,=\,1 $t_{2g}^2$ ground state to low-spin $S$\,=\,0 $t_{2g}^2$ states at 1.07\,eV and \,1.28\,eV
(these are allowed due to the strong spin-orbit coupling of the core hole), and further excitations above 1.8\,eV.\@ Our central results are the observations of a finite dispersion of the low-energy orbital excitations at 0.1\,-\,0.2\,eV and of an additional weak peak at 0.4\,eV for x-ray energies tuned to the O $K$ edge. The combination of RIXS and optical data\cite{Benckiser2008} gives strong evidence for the interpretation of the 0.4\,eV peak as a two-orbiton excitation.

The paper is organized as follows. Experimental details are given in Sec.~\ref{experiment}, followed in Sec.~\ref{struc} by a short description of the crystal structure, the spin/orbital-ordered phases, and the electronic structure. Section~\ref{results} describes our results. In Sec.~\ref{SecXAS} we briefly discuss x-ray absorption spectra. The RIXS features observed at the V $L_3$ edge are assigned to the different orbital excitations in Sec.~\ref{assign}. In Sec.~\ref{lowenergy} we concentrate on the low-energy orbital excitations observed at 0.1\,-\,0.2\,eV, focussing on the observation of a finite dispersion and on the competition between superexchange interactions and the crystal field. Section~\ref{Kedge} addresses the RIXS data measured at the O $K$ edge. Conclusions are given in Sec.~\ref{conclusion}. In the appendix, we discuss optical data on crystal-field excitations of V$^{3+}$ ions in DySc$_{0.9}$V$_{0.1}$O$_3$.

\section{Experimental Details} \label{experiment}

\begin{figure}[tb]
\includegraphics[width=0.99\columnwidth]{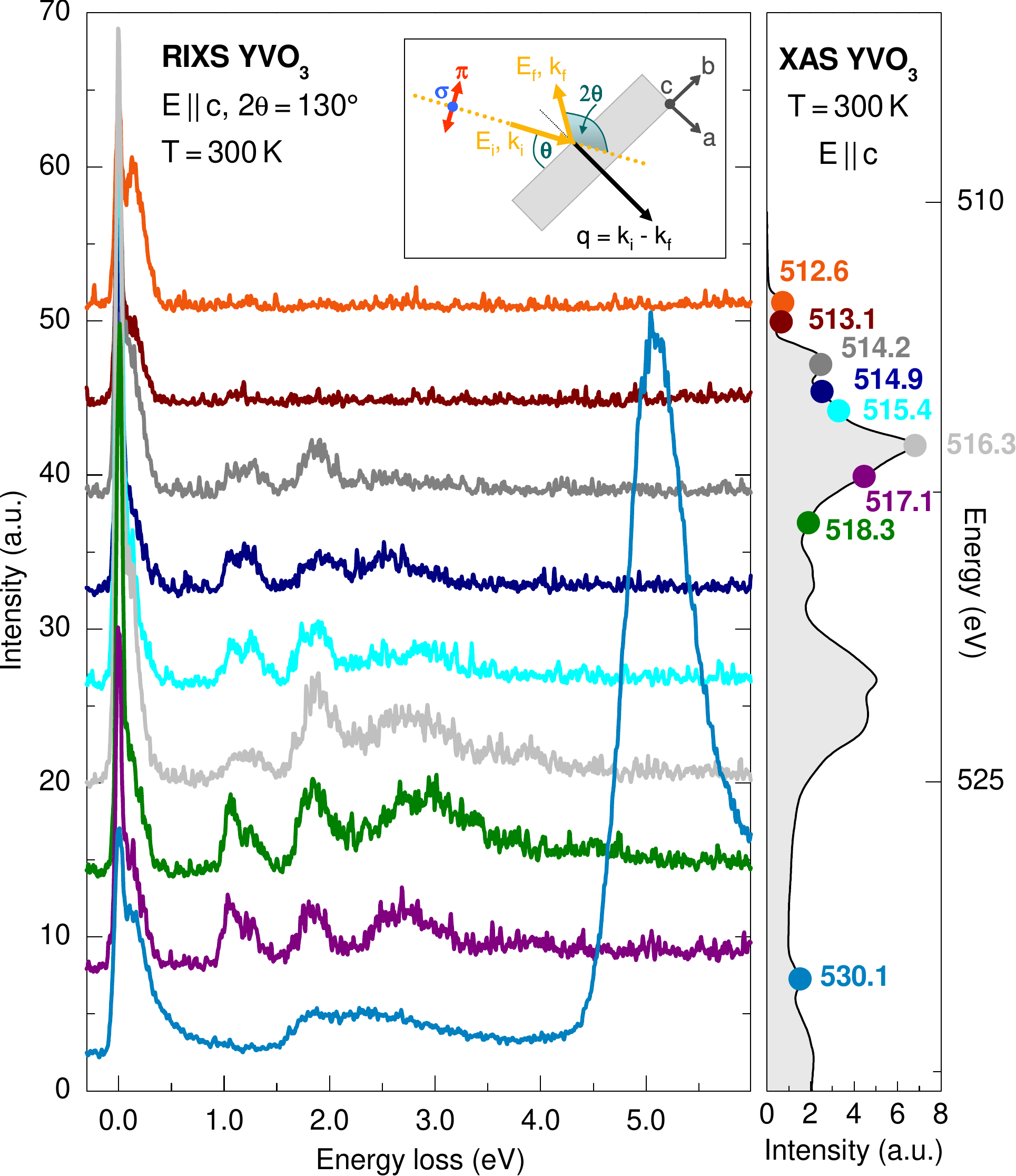}
\caption{RIXS overview spectra (lower statistics, accumulated over 30\,min each) at $T$\,=\,300\,K for nine different incident energies across the V $L_3$ edge (512\,-\,519\,eV) and at the O $K$ edge ($\geq$\,530\,eV). The x-ray absorption spectrum (XAS) measured in total electron yield is shown in the right panel, with color-coded points indicating the excitation energies $E_{\rm in}$ used for the RIXS spectra. A sketch of the scattering geometry is shown in the inset. All spectra were measured with $E$\,$\parallel$\,$c$. The RIXS spectra were measured at a scattering angle of $2\theta$\,=\,130$^\circ$. The inelastic peaks with energy transfers up to 4\,eV correspond to orbital excitations. The steep increase of the RIXS signal at 4.5\,eV for $E_{\rm in}$\,=\,530.1\,eV at the O \textit{K} edge is in good agreement with the onset of charge-transfer excitations observed in the optical conductivity.\cite{Miyasaka2002,Mossanek2009,Reul12} However, the large intensity, exceeding the intensity of the elastic peak, points towards x-ray fluorescence emission, possibly superimposed by the charge-transfer excitations mentioned above.}
\label{Fig1}
\end{figure}

Single crystals of YVO$_3$ were grown by the traveling-solvent floating zone method.\cite{Blake2002} Prior to the experiments the samples have been freshly cleaved in air. The measurements were performed using the soft x-ray spectro\-meter SAXES at the ADRESS beam line at the Swiss Light Source, Paul Scherrer Institut, Villigen, Switzerland, which has been designed especially for high-resolution RIXS.\cite{Strocov2010,Ghir06a} The RIXS data were measured with a combined energy resolution
(monochromator and spectrometer) of $\Delta E$\,=\,60\,meV at the V $L_3$ edge and $\Delta E$\,=\,70\,meV at the O $K$ edge for scattering angles of $2\theta$\,=\,90$^{\circ}$ and 130$^{\circ}$ (see sketch in Fig.~\ref{Fig1}). All measurements were performed in specular geometry to keep the transferred momentum $q$ parallel to the crystallographic $a$ axis. For this purpose, the sample was oriented with the $c$ axis perpendicular to the scattering plane and the incident x-rays
were linearly polarized with $E$\,$\parallel$\,$c$ ($\sigma$ polarization). Compared to $\pi$ polarization parallel to the scattering plane, $\sigma$ polarization has the advantage that the orientation of the electric field with respect to the crystallographic axes is independent of the scattering angle. The polarization of the scattered light has not been determined. The attempt to measure a second sample with a different orientation to obtain data for $q \! \parallel \! c$ failed due to experimental problems with this sample.

The incident energy has been varied across the V \textit{L}$_{3}$ and O \textit{K} edges. Before the RIXS measurements, room-temperature x-ray absorption (XAS) spectra were measured \textit{in situ} in total electron yield (TEY) mode (see right panel in Fig.\ \ref{Fig1}). Below about 100\,K, the resistivity of the samples is very high, thus charging effects constricted the measurement of the drain current. Therefore, we additionally measured XAS spectra in total fluorescence yield (TFY) mode (inset in Fig.~\ref{Fig2}). Note that the absolute energy scale has been calibrated by a comparison with the XAS data of polycrystalline YVO$_3$ and of V$_2$O$_3$.\cite{Pen1999,Park2000}

\section{Crystal Structure, Magnetic and Orbital Order, and Crystal-Field Levels}\label{struc}

The pseudo-perovskite compound YVO$_3$ shows several phase transitions as a function of temperature. In the room-temperature phase an orthorhombic structure (\textit{Pbnm}) is adopted, while for $77\,K < T < 200\,K$ a monoclinic structure (\textit{P2$_1$/b}) with long-range orbital order was observed\cite{Blake2002,Reehuis2006}($G$ type, possibly with an admixture of $C$ type, see discussion in Ref.\ \onlinecite{Benckiser2008}). Antiferromagnetic order of
$C$ type was found below $T_N$\,=\,116\,K.\@ Below 77\,K, the structure becomes orthorhombic (\textit{Pbnm}) again, accompanied by a change to $C$-type orbital and $G$-type spin order.\cite{Blake2002,Reehuis2006} As far as the electronic structure is concerned, YVO$_3$ is a Mott insulator with two localized electrons in the $3d$ shell of each V$^{3+}$ ion. A crystal field of predominantly octahedral symmetry gives rise to a splitting of the $3d$ levels into a lower-lying $t_{2g}$ level and
an $e_g$ level. These are further split by the deviations from cubic symmetry.\cite{deRay07,Solovyev06,Solovyev08,Otsuka06,Benckiser2008} In cubic approximation, the lowest-lying two-electron state shows $^{3}T_{1}$ symmetry ($t_{2g}^2$, $S$\,=\,1), the lowest excited states are of $^{1}T_{2}$ and $^{1}E$ symmetry ($t_{2g}^2$, $S$\,=\,0) and are expected above 1\,eV (see Sec.\ \ref{assign}). At still higher energies, states with the following symmetries are expected: $^{3}T_{2}$ ($t_{2g}^1 e_{g}^1$, $S$\,=\,1), $^{1}\!A_{1}$ ($t_{2g}^2$, $S$\,=\,0), as well as $^{3}T_{1}(P)$, $^{1}T_{2}$, and $^{1}T_{1}$ ($t_{2g}^1 e_{g}^1$, $S$\,=\,0 or 1).\cite{Sugano}
The cubic approximation is sufficient for the overall assignment of all RIXS peaks observed above 1\,eV at the V $L_3$ edge. However, for the discussion of orbital excitations below about 0.5\,eV we have to take into account deviations from cubic symmetry (see Secs.\ \ref{assign} and \ref{lowenergy}).

\section{Results and Discussion} \label{results}

\subsection{XAS spectra} \label{SecXAS}

\begin{figure}[tb]
\includegraphics[width=0.99\columnwidth]{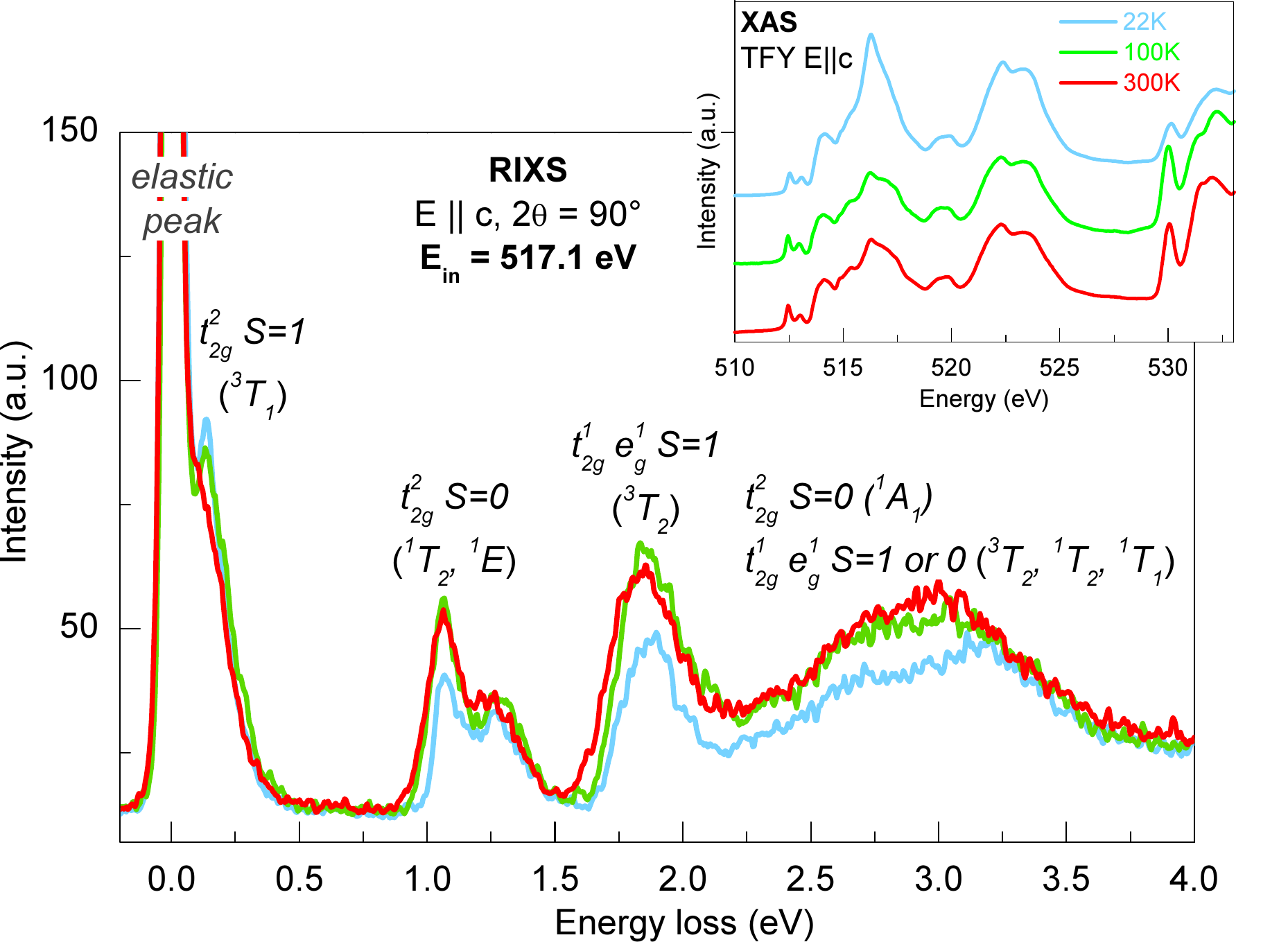}
\caption{RIXS signal of YVO$_3$ for $E$\,$\parallel$\,$c$ at 22\,K, 100\,K, and 300\,K for $E_{\rm in}$\,=\,517.1\,eV and 2$\theta$=90$^{\circ}$. These data sets have been accumulated over 3\,h each. Inset: Temperature-dependent XAS spectra of YVO$_3$ for polarization $E$\,$\parallel$\,$c$ at 22\,K, 100\,K, and 300\,K recorded in total fluorescence yield mode.}
\label{Fig2}
\end{figure}

XAS spectra have been measured in total fluorescence yield at 22\,K, 100\,K, and 300\,K, i.e., in all three crystallographic phases (inset of Fig.~\ref{Fig2}). The absorption bands peaking at about 516\,eV, 522\,eV, and 534\,eV correspond to the V $L_3$, V $L_2$, and O $K$ edges, respectively, i.e., transitions from the spin-orbit split V $2p_{3/2}$ and V $2p_{1/2}$ core states to empty V $3d$ states, and from O $1s$ to O $2p$ states, respectively. Empty states within the $2p$ level of oxygen arise due to hybridization. In particular, the lowest peak at the O $K$ edge at 530.1\,eV is attributed to states of mixed O $2p$ - V $3d$ character.\cite{Pen1999,Mossanek2009} A detailed analysis of the x-ray absorption spectra of YVO$_3$ can be found in Refs.~\onlinecite{HollmannDiss} and \onlinecite{Pen1999}.

\begin{figure}[tb]
\includegraphics[width=0.9\columnwidth]{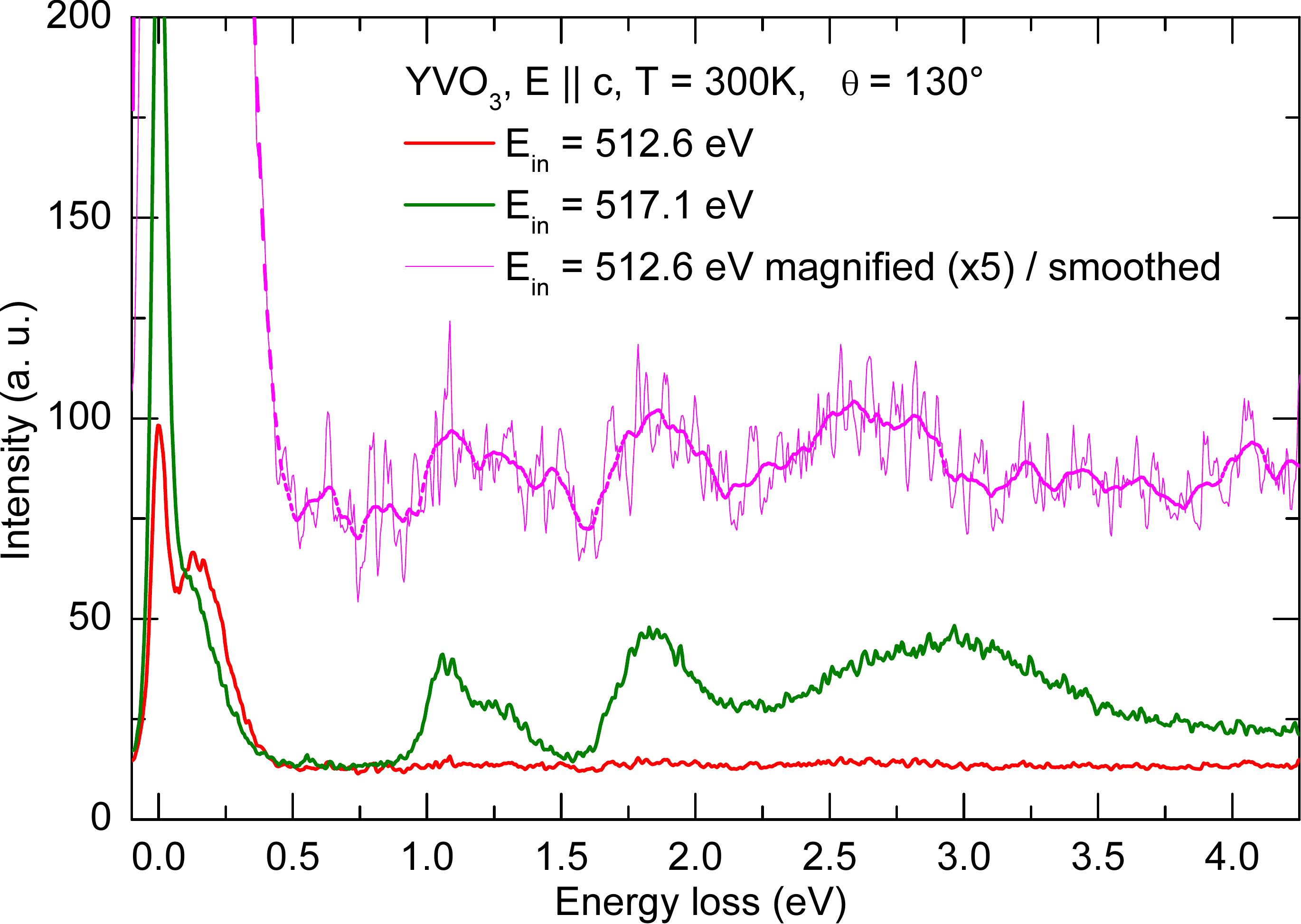}
\caption{Comparison of the RIXS signal for $E_{\rm in}$\,=\,512.6\,eV (red) and 517.1\,eV (green). The thin magenta curve shows the same data for $E_{\rm n}$\,=\,512.6\,eV on a larger scale (thick line: smoothed). The strong high-energy RIXS features observed for $E_{\rm in}$\,=\,517.1\,eV can also be resolved for $E_{\rm in}$\,=\,512.6\,eV but with strongly reduced spectral weight.}
\label{Fig3}
\end{figure}

\begin{table}[tb]
\begin{tabular}{l c c c c}
\hline\smallskip
    & \footnotesize{$t_{2g}^2$\,($S$=1)} & \footnotesize{$t_{2g}^2$\,($S$=0)} & \footnotesize{$t_{2g}^1 e_g^1$\,($S$=1)} & \footnotesize{$t_{2g}^1 e_g^1$\,($S$=0, 1) }\\
    & & & & \footnotesize{and $t_{2g}^2$\,($S$=0)}\\
\hline\hline
\multicolumn{5}{l}{\textbf{RIXS 2$\theta$=90$^{\circ}$}}\\
\footnotesize{300K}   & \footnotesize{(0.10\,$\&$\,0.20)$^{\dagger}$} & \footnotesize{1.07\,$\&$\,1.25}       & \footnotesize{1.84}   &  \footnotesize{2.2 - 3.5}\\
\footnotesize{100K}   & \footnotesize{(0.11\,$\&$\,0.20)$^{\dagger}$} & \footnotesize{1.07\,$\&$\,1.28}       & \footnotesize{1.86}   &  \footnotesize{2.2 - 3.5}\\
\footnotesize{22K}    & \footnotesize{(0.12\,$\&$\,0.20)$^{\dagger}$} & \footnotesize{1.07\,$\&$\,1.27}       & \footnotesize{1.88}   &  \footnotesize{2.2 - 3.5}\\
\hline
\multicolumn{5}{l}{\textbf{RIXS 2$\theta$=130$^{\circ}$}}\\
\footnotesize{300K}   & \footnotesize{(0.10\,$\&$\,0.20)$^{\dagger}$}& \footnotesize{1.07\,$\&$\,1.25}        & \footnotesize{1.84}   &  \footnotesize{2.2 - 3.5}\\
\footnotesize{100K}   & \footnotesize{(0.13\,$\&$\,0.22)$^{\dagger}$}& \footnotesize{1.07\,$\&$\,1.28}       & \footnotesize{1.86}   &  \footnotesize{2.2 - 3.5}\\
\hline
\multicolumn{5}{l}{\textbf{Optics}}\\
\footnotesize{300K}   & \footnotesize{0.12\,-\,0.22$^\ddag$}            & \footnotesize{1.1-1.2}   & \footnotesize{1.84-1.87$^{*\ddag}$}  & \footnotesize{$>$\,2.2$^{*\ddag}$}  \\
\footnotesize{100K}   & \footnotesize{0.12\,$\&$\,0.22$^\ddag$}     & \footnotesize{1.1-1.3}   &     &               \\
\footnotesize{22K}    & \footnotesize{0.12\,$\&$\,0.21$^\ddag$}     & \footnotesize{1.1-1.3}   &     &               \\
\hline
\multicolumn{5}{l}{\textbf{Theory \footnotesize{(Ref.\ \onlinecite{deRay07})}}}\\
\footnotesize{300K}        & \footnotesize{0.08\,$\&$\,0.21}                 & & & \\
\footnotesize{100K (V1)}   & \footnotesize{0.08\,$\&$\,0.24}                 & & &\\
\footnotesize{100K (V2)}   & \footnotesize{0.06\,$\&$\,0.20}                & & & \\
\footnotesize{65K}         & \footnotesize{0.08\,$\&$\,0.20}                 & & & \\
\hline
\end{tabular}
\caption{Comparison of energies (in eV) of the orbital excitations in YVO$_3$ from a) RIXS (this work, $2\theta$\,=\,90$^{\circ}$), b) RIXS (this work, $2\theta$\,=\,130$^{\circ}$), c) optical conductivity (Ref.~\onlinecite{Benckiser2008}), and d) LDA calculations (Ref.~\onlinecite{deRay07}), where V1 and V2 refer to the two different V sites present in the intermediate phase. The values labeled by ${\dagger}$ were obtained by fits of two Gaussians (see Fig.~\ref{Fig5}). Values labeled by ${*}$ were obtained from absorption data of DyV$_{0.1}$Sc$_{0.9}$O$_3$ (for details see the Appendix and Ref.~\onlinecite{BenckDiss}.) For the values labeled by $\ddag$, a phonon shift of $E_{\rm ph}$\,=\,50\,-\,80\,meV has been subtracted from the peak values observed in optics.}\label{Tab1}
\end{table}

\subsection{Assignment of RIXS peaks (V $L_3$ edge)} \label{assign}

Figure~\ref{Fig1} shows RIXS spectra at room temperature as a function of the transferred energy for different incident energies $E_{\rm in}$ which are indicated by color-coded dots in the XAS data (right panel). Each spectrum has been accumulated over 30 minutes. Long-time scans with better statistics, accumulated over 3 hours are given in Fig.\ \ref{Fig2} for $E_{\rm in}$\,=\,517.1\,eV.\@ At the V $L_3$ edge, we observe a series of inelastic peaks with energy transfers of about 0.10\,-\,0.20\,eV, 1.07\,eV, 1.25\,eV, and 1.84\,eV, and a broad band of overlapping features between 2.2\,eV and 3.5\,eV at 300\,K.\@ The spectral weights of these peaks strongly depend
on $E_{\rm in}$ but the peak energies do not (see Figs.~\ref{Fig1} and \ref{Fig3}), clearly demonstrating the Raman character of these modes. With increasing $E_{\rm in}$, the spectral weight of the higher-lying bands increases, which reflects the change of the intermediate state and the corresponding change of the resonance conditions.

The main contribution to the RIXS process at the V $L_3$ edge can be denoted by $2p^6 3d^2 \rightarrow 2p^5 3d^{3} \rightarrow 2p^6 [3d^2; 3d^{2*}]$.\cite{Ghir05,Ament11} In the first step a V $2p_{3/2}$ core electron is excited by a photon with $E_{\rm in}$ to the open $3d$ shell. In the second step an electron relaxes to the $2p$ shell,
leaving the system either in the initial ground state ($3d^2$, elastic peak), or in a low-energy excited state ($3d^{2*}$, inelastic peaks). For energies smaller than the Mott-Hubbard gap of about 1.6\,eV,\cite{Benckiser2008} this low-energy excitation can be a phonon, a magnon, or an orbital excitation. In general, also combined excitations such as an orbital excitation with phonon (or magnon) sidebands are allowed. In YVO$_3$ phonon energies are smaller than 0.09\,eV,\cite{Tsvetkov2004,Miyasaka2005,Miyasaka2006,Sugai2006,Jandl2010} and magnon energies do not exceed 0.04\,eV,\cite{Ulrich2003} thus we attribute the observed RIXS features to orbital excitations. Indeed, the excitation energies observed in RIXS are in excellent agreement with the values derived for the orbital excitations
by optical absorption measurements.\cite{Benckiser2008} For the sake of completeness we note that these orbital excitations may show phonon (or magnon) sidebands,
which may be important for the discussion of the line width (see Sec.~\ref{linewidth}). YVO$_3$ shows inversion symmetry on the V sites. Thus orbital (or $d$-$d$) excitations are forbidden in optics by parity and do not directly contribute to the optical absorption spectrum due to the dipole selection rule. A weak contribution to the optical spectrum arises by the simultaneous excitation of a phonon. In oxides, the O phonon modes of 50\,-\,80\,meV are most effective. In contrast, an orbital excitation in RIXS corresponds to two subsequent dipole-allowed transitions, thus orbital excitations contribute directly to the RIXS signal. When comparing the results of the two spectroscopies for the energies of the spin-conserving excitations from the $t_{2g}^2$ $S$\,=\,1 ground state to $S$\,=\,1 final states (see Tab.~\ref{Tab1}),
the phonon shift occurring in optical absorption data has to be taken into account.

At the V $L_3$ edge, the overall RIXS spectrum (see Fig.~\ref{Fig2}) can be well described in a local crystal-field scenario. For a $t_{2g}^2$ high-spin $S$\,=\,1 ground state, we expect the following orbital excitations with increasing energy: spin-conserving intra-$t_{2g}$ excitations at low energies; spin-flip intra-$t_{2g}$ excitations to a low-spin $S$\,=\,0 final state at about $2J_H$, where $J_H\! \approx\! 0.6-0.7$\,eV denotes Hund's coupling;\cite{Zaanen1990,Mizokawa96} and $t_{2g}^2 \rightarrow t_{2g}^1 e_g^1$ excitations at higher energies (see Tab.~\ref{Tab1}). In the optical data,\cite{Benckiser2008} the spin-conserving intra-$t_{2g}$ excitations
have been identified with a feature at about 0.20\,-\,0.26\,eV.\@ Subtracting a phonon shift of 50\,-\,80\,meV, this yields 0.12\,-\,0.21\,eV for the orbital excitations, in excellent agreement with the RIXS peak at about 0.1\,-\,0.2\,eV.\@ These values also agree with various theoretical results predicting the intra-$t_{2g}$ transitions in the range of 0.06\,-\,0.24\,eV.\cite{deRay07,Solovyev06,Solovyev08,Otsuka06} Based on Raman scattering data, the observation of orbitons has been claimed in $R$VO$_3$ at lower excitation energies of 43\,meV and 62\,meV (Refs.~\onlinecite{Miyasaka2005,Miyasaka2006}) and at 45\,meV and 84\,meV.\cite{Sugai2006} However, a more recent study\cite{Jandl2010} suggests that these peaks have to be interpreted as multiphonons, in agreement with our results. A more detailed analysis of the low-energy RIXS feature is given in Sec.~\ref{lowenergy}.

Intra-$t_{2g}$ excitations from the high-spin $S$\,=\,1 ground state to a low-spin $S$\,=\,0 final state are expected at about $2J_H \! \approx \! 1.2\,-\,1.4$\,eV.\cite{Zaanen1990,Mizokawa96} In optics, these excitations give rise to a rich structure with dominant peaks at about 1.1\,-\,1.3\,eV,\cite{Benckiser2008} again in excellent agreement with the RIXS data. The value of $J_H$ is hardly screened in a solid, thus very similar energies are observed for $V^{3+}$ ions in different compounds, e.g., V-doped $\alpha$-Al$_2$O$_3$.\cite{Dongping97} In the intermediate state of the RIXS process at the V $L_3$ edge, spin is not a good quantum number due to the large spin-orbit splitting between V $2p_{3/2}$ and V $2p_{1/2}$ core-hole states. Therefore, the spin-selection rule does not apply, giving rise to a large spectral weight of such high-spin to low-spin excitations.\cite{deGroot1998,Ghiringhelli2009} This is corroborated by the absence of RIXS peaks between 1.1 and 1.3\,eV in the data obtained at the O $K$ edge (see Fig.~\ref{Fig1}). The $1s$ core hole in the corresponding intermediate state does not show an orbital moment, thus the spin-orbit mechanism is \lq{}switched off\rq{} and the spin-selection rule applies. Also in optics, the spin-flip excitations are suppressed by the spin selection rule, but they become weakly allowed by spin-orbit coupling or by the simultaneous excitation of a magnon.\cite{Sugano,Figgis} Therefore, it is not necessary to consider a phonon shift for the comparison of optics and RIXS in the case of spin-flip excitations.

The RIXS peak observed at 1.84\,eV at 300\,K (see Fig.~\ref{Fig2}) corresponds to excitations from the $t_{2g}$ level into the $e_g$ level, $t_{2g}^2 \rightarrow t_{2g}^1 e_g^1$, i.e., to excitations into $^3T_2$ in cubic approximation. A value of roughly 2\,eV is typical for a V$^{3+}$ ion in an oxygen octahedron.\cite{Sugano,Figgis,Dongping97,Ishii02} Again, RIXS and optical results are in excellent agreement with each other (see Appendix). A detailed assignment of the higher-lying excitations is hindered by the fact that the different contributions strongly overlap in the RIXS data. At 300\,K, we observe a peak at about 3.0\,eV with shoulders at about 2.3\,eV and 2.7\,eV.\@ For the $3d^2$ configuration in a ligand field of intermediate strength, the Tanabe-Sugano diagram\cite{Sugano,Figgis} predicts in cubic approximation that the energy of the spin-flip excitation $^1\!A_1$ is about 2.1\,-\,2.2 times the energy of the $^1T_2\,/\,^1\!E$ band observed at 1.07\,-\,1.25\,eV.\@ Thus we expect the $^1\!A_1$ peak roughly at 2.2\,-\,2.8\,eV, i.e., in the range of the shoulders at 2.3\,eV and 2.7\,eV.\@ Above the $^1\!A_1$ peak, the Tanabe-Sugano diagram predicts excitations to $^3T_1(P)$, $^1T_2$, and $^1T_1$. As noted above for the peaks at 1.07\,-\,1.25\,eV, the orbital excitation energies observed here are similar to the ones reported for V$^{3+}$ ions doped into $\alpha$-Al$_2$O$_3$, in which the $^1\!A_1$ and $^3T_1(P)$ excitations lie at 2.6\,eV and 3.1\,eV, respectively.\cite{Dongping97} For the data measured at the V $L_3$ edge we conclude that the RIXS spectrum above about 1\,eV can be well described in terms of local crystal-field excitations.

\subsection{Low-energy orbital excitations at 0.1\,-\,0.2\,eV} \label{lowenergy}

\begin{figure}[b]
\center\includegraphics[width=0.9\columnwidth]{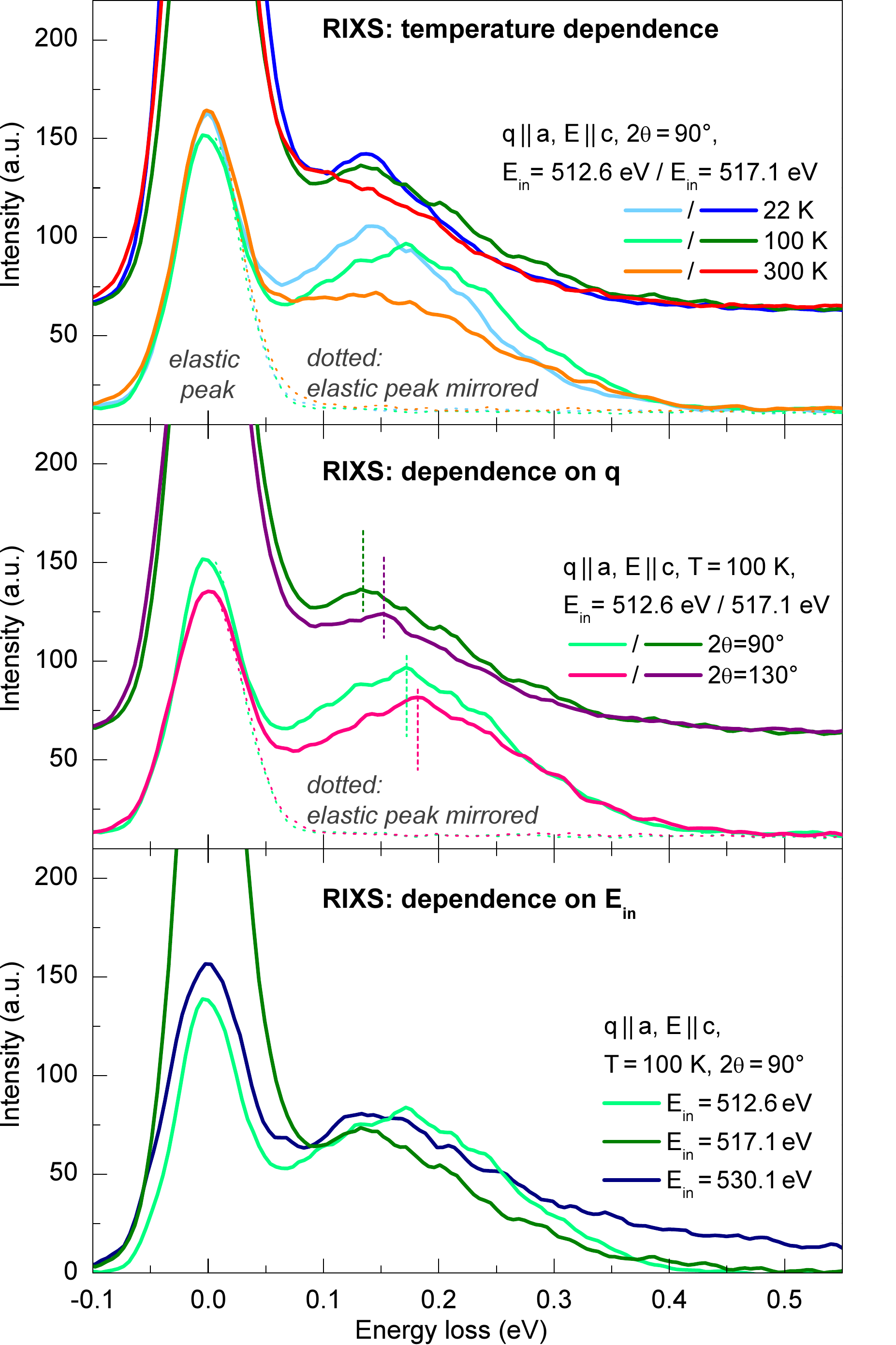}
\caption{Low-energy RIXS spectra of YVO$_3$ at the V $L_3$ edge. Top: Temperature dependence of the intra-$t_{2g}$ excitations with $2\theta$\,=\,90$^{\circ}$ and $E_{\rm in}$\,=\,512.6\,eV or 517.1\,eV at 22\,K, 100\,K, and 300\,K.\@ Middle: Dependence on the scattering angle $2\theta$ with $2\theta$\,=\,90$^{\circ}$ and 130$^{\circ}$ for $T$\,=\,100\,K and different values of $E_{\rm in}$. Bottom: Dependence on $E_{\rm in}$. Data for $E_{\rm in}$\,=\,517.1\,eV are offset for clarity in the top and middle panels. All spectra were accumulated over 3 hours with $E$\,$\parallel$\,$c$ and $q$\,$\parallel$\,$a$. To estimate the contribution from the elastic peak with a full width at half maximum of 60\,meV, the spectra have been mirrored around zero (dotted lines).}
\label{Fig4}
\end{figure}

Our main goal is to clarify whether a local crystal-field scenario fully describes the orbital excitation spectrum, or whether superexchange plays a significant role. A thorough quantitative analysis has to consider both the crystal field and superexchange interactions simultaneously. However, this complex problem has hardly been addressed thus far.\cite{Brink01,Schmidt07,Krivenko12,Horsch2008} Therefore we focus on the two limiting cases, either a dominant crystal field or dominant superexchange interactions. Both, superexchange and the non-cubic crystal field lift the threefold orbital degeneracy of the cubic $^3T_1$ ground state, giving rise to low-energy intra-$t_{2g}$ excitations. We concentrate on these excitations below 0.5\,eV, where the energy scale of superexchange interactions may become comparable to the excitation energy itself. Figure~\ref{Fig4} depicts the dependence of the low-energy RIXS spectra on temperature $T$ (top panel) and crystal momentum $q$ (middle panel)
for incident energies of $E_{\rm in}\,=\,512.6$\,eV and 517.1\,eV, as well as the dependence on $E_{\rm in}$ (bottom panel).

\begin{figure*}[t]
\includegraphics[width=0.8\linewidth]{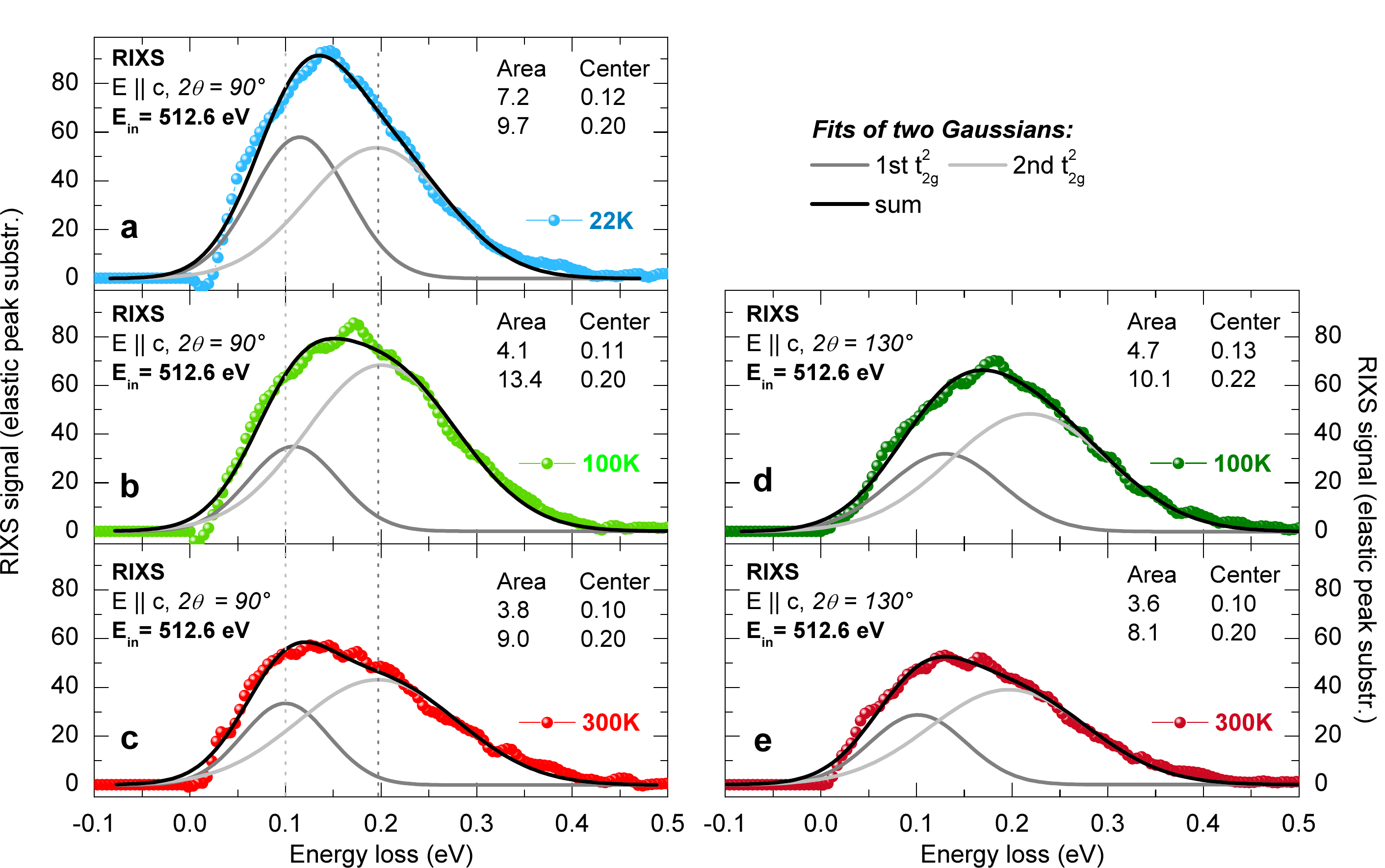}
\caption{RIXS spectra of the intra-$t_{2g}$ excitations for different temperatures (top: 22\,K; middle: 100\,K; bottom: 300\,K) and different scattering angles
(left: $2\theta$\,=\,90$^{\circ}$; right: 130$^{\circ}$) for $E_{\rm in}$\,=\,512.6\,eV.\@ The elastic line has been subtracted by mirroring it around zero (see Fig.~\ref{Fig4}). In order to account for the two-peak structure, we fitted two Gaussians to the data (solid lines). The peak positions of both peaks show shifts of less than 40\,meV as a function of temperature. Most of the observed changes of the line shape can be explained by the variation of spectral weight, i.e., of the peak areas. At 100\,K, both Gaussian peaks show a shift of about 20\,meV as a function of the scattering angle $2\theta$.}
\label{Fig5}
\end{figure*}

\subsubsection{Line width, peak energy, and number of peaks}\label{linewidth}

The overall line width of about 0.2\,eV of the features at 0.1\,-\,0.2\,eV is certainly larger than the experimental resolution of 60\,meV.\@ We identify two main reasons for the large line width: (i) a multi-peak structure, i.e.\ the experimental feature is composed of more than one peak, and (ii) the coupling to phonons or magnons, giving rise to incoherent parts of the orbital spectra.

In a crystal-field scenario, the coupling of the orbital degrees of freedom to phonons gives rise to a vibronic character of the elementary excitations. The spectrum is composed of a series of phonon \lq{}sidebands\rq{}. In a solid, these sidebands are typically not resolved but form a broad excitation continuum.\cite{Figgis,HendersonImbusch} Concerning point (i), a multi-peak structure arises both from the splitting of the orbitally threefold degenerate $^3T_1$ state in a non-cubic environment and from the existence of two different V sites within the unit cell in the intermediate phase at 100\,K. In total, this yields four different excitation energies in a crystal-field scenario. However, results based on LDA\cite{deRay07} (local density approximation) and first principles\cite{Solovyev08} indicate
that the crystal-field splitting is similar for the two V sites and that the excitation spectrum can be grouped into two bands (at 0.06\,-\,0.08\,eV and 0.20\,-\,0.24\,eV according to Ref.~\onlinecite{deRay07}, see Tab.~\ref{Tab1}, or at 0.10\,-\,0.13\,eV and 0.20\,eV according to Ref.~\onlinecite{Solovyev08}) which reflect the splitting of the cubic $^3T_1$ state.

In a superexchange scenario, one also expects a multi-peak structure, namely two different one-orbiton modes. For the intermediate state at $T$\,=\,100\,K with $C$-type spin order and $G$-type orbital order, one mode is expected to show a dispersion for $q\! \parallel \! c$ (see below) whereas the other one remains dispersionless.\cite{Oles2007,Ishihara2004} A model based only on superexchange neglects the coupling to phonons and thus neglects the vibronic broadening. However, orbital excitations are also coupled to magnons via the common superexchange processes, giving rise to broad incoherent parts of the orbital spectra.\cite{Wohlfeld11,Wohlfeld12} Even broader features are expected for two-orbiton excitations, which will be discussed in Sec.~\ref{Kedge}.

In the RIXS data, it is not possible to directly distinguish different contributions to the broad feature at 0.1\,-\,0.2\,eV. However, there is an interesting dependence on $E_{\rm in}$ at the V $L_3$ edge (see Fig.~\ref{Fig4}). The RIXS peak maximum is at 0.13\,-\,0.14\,eV for $E_{\rm in}$\,=\,517.1\,eV, about 30\,-\,40\,meV lower than for $E_{\rm in}$\,=\,512.6\,eV. This is valid both for $2\theta$\,=\,90$^{\circ}$ and 130$^{\circ}$. At the same time, the inelastic peaks extend to about 0.4\,eV for all data obtained at the V $L_3$ edge, irrespective of the shift of the peak maximum. This can be rationalized by assuming that the RIXS feature consists of two different peaks with spectral weights that depend on $E_{\rm in}$, similar to the overall behavior depicted in Fig.~\ref{Fig1}. A strong dependence of the relative spectral weight of orbital excitations was observed also in RIXS on TiOCl.\cite{Glawion11} We emphasize that it is possible to describe our RIXS data by two orbital excitations at 0.10\,-\,0.13\,eV and 0.20\,-\,0.22\,eV with Gaussian line shapes under the assumption that the spectral weights of these two excitations depend on $E_{\rm in}$, temperature, and momentum (see Fig.~\ref{Fig5}).

A two-peak scenario is supported not only by the LDA and first-principles results mentioned above\cite{deRay07,Solovyev08} but also by the optical conductivity, showing these excitations with a higher energy resolution at about 0.20\,eV and 0.26\,eV.\cite{Benckiser2008} Subtracting a phonon shift of 50\,-\,80\,meV from the optical results yields orbital excitation energies of 0.12\,-\,0.15\,eV and 0.18\,-\,0.21\,eV, in agreement with our analysis of the RIXS data (see Tab.~\ref{Tab1}).

The orbital excitation energies observed in RIXS agree with the expectations of a crystal-field scenario but are hard to reconcile with a \textit{pure} superexchange scenario. The energy scale for superexchange interactions is given by $J$\,=\,$4t^2/U$, where $t$ denotes the nearest-neighbor V-V hopping parameter and $U$ the on-site Coulomb repulsion $U$\,$\approx$\,5\,eV. The value of $J$ can been determined from, e.g., the spin-wave dispersion observed in inelastic neutron scattering.\cite{Ulrich2003} Note that $J$ sets the overall energy scale for a spin-orbital superexchange model but that the effective spin-exchange coupling constants $J_{ab}^{\rm spin}$ and $J_{c}^{\rm spin}$ are smaller than $J$, they depend on the orbital correlations and thus on the ground state (see, e.g., Fig.~4 in Ref.~\onlinecite{Oles2007}). Accordingly, different estimates have been given\cite{Ulrich2003,Oles2007,Horsch2008} for the size of $J$ (0.02\,-\,0.04\,eV). For $q\!\parallel \!c$, the dispersion of orbital excitations in the intermediate phase between 77\,K and 116\,K with $C$-type spin order and $G$-type orbital order is given by\cite{Oles2007}
\begin{equation}
\omega_c(q) = J \sqrt{\Delta^2 + \left(1-3\frac{J_H}{U}\right)^{-2}\sin^2{qc}}
\label{dispersion}
\end{equation}
with $J_H/U$\,$\approx$\,0.13 and a gap of $\Delta\cdot J$. The parameter $\Delta$ depends on $J_H/U$ and on the crystal field and is explicitly given in Ref.~\onlinecite{Oles2007}. Neglecting the crystal field, one finds $\Delta$\,$\approx$\,1 and a maximum energy of orbital excitations of about 0.08\,eV.\cite{Oles2007} For $J$\,=\,40\,meV, the different estimates of the crystal-field contribution yield orbital excitations with a gap of about 0.02\,-\,0.10\,eV and a bandwidth of about 0.02\,-\,0.05\,eV for YVO$_3$. In comparison, the observed excitation energies of 0.1\,-\,0.2\,eV are rather large and clearly indicate a significant contribution of the crystal field. We like to add that a superexchange scenario predicts a significantly reduced dispersion for the low-temperature phase below 77\,K with $G$-type spin order and $C$-type orbital order.\cite{Ishihara2004,Oles2007} Also for the low-temperature phase, a finite dispersion is only expected for $q\!\parallel \!c$.

At first sight, a superexchange model seems to suggest an alternative scenario in which the lower peak at 0.1\,eV corresponds to single-orbiton excitations and the upper one at 0.2\,eV to a two-orbiton contribution, as discussed for the titanates.\cite{Ulrich2009,Ament2010} Here, one has to discuss the relationship between a single local orbital flip on a given transition-metal site (i.e., the RIXS final state at the V $L_3$ edge) and the number of orbitons that are excited. In a crystal-field scenario, flipping an orbital is equivalent to a single local crystal-field excitation. Two orbital flips on adjacent sites then correspond to two crystal-field excitations.
This simple correspondence breaks down in a superexchange scenario, reflecting many-body physics and the quasiparticle character of the excitations. A local orbital flip has to be translated into the eigenstates of the bulk, i.e., into a superposition of one- and multi-orbiton excitations, similar to one- and multi-magnon contributions in case of a local spin flip.\cite{Haverkort2010} However, a strong two-orbiton contribution at the transition-metal $L$ edge is only expected in case of strong orbital fluctuations.\cite{Ament2010} For YVO$_3$, strong orbital fluctuations have been ruled out recently by optical data for the Mott-Hubbard excitations.\cite{Reul12} Another argument against a two-orbiton interpretation of the peak at 0.2\,eV is the observation of a two-orbiton peak at 0.4\,eV in RIXS at the O $K$ edge (see Sec.~\ref{Kedge}) and in the optical conductivity.\cite{Benckiser2008} Finally, the temperature dependence of the peak energies is hard to explain in a scenario with dominant superexchange interactions, as discussed in the next paragraph. Therefore, a two-orbiton interpretation of the RIXS feature at 0.2\,eV can be ruled out.

\subsubsection{Temperature dependence}

The RIXS data for different temperatures resolve the dependence of the intra-$t_{2g}$ excitations on the crystal structure in the different phases of YVO$_3$ very well (see top panel of Fig.~\ref{Fig4}). From a phenomenological point of view, we may neglect the two-peak structure discussed above and consider solely the maximum of the experimental peak or its first moment, i.e., the center of mass of the inelastic spectrum. Both the maximum and the first moment show shifts of the order of 10\,-\,40\,meV as a function of temperature. Similar shifts are obtained both from the fits using two Gaussians for the two-peak structure (see Fig.~\ref{Fig5} and Tab.~\ref{Tab1}) and from the optical data.\cite{Benckiser2008} A shift of the excitation energy can be explained in a crystal-field scenario by a change of the crystal structure and thus of the crystal field, and in a superexchange-based scenario by a change of the orbital order.\cite{Oles2007,Ishihara2004} However, the orbital order changes dramatically between the different ordered phases and the orbitally disordered phase above 200\,K, and orbital fluctuations are only weak above 200\,K.\cite{Reul12} Accordingly, spin-orbital superexchange models predict a significant change of the excitation energies across the phase transitions between the different phases.\cite{Ishihara2004,Oles2007} In contrast, the observed shifts of 10\,-\,40\,meV are much smaller than the excitation energy of 0.1\,-\,0.2\,eV. This clearly indicates that the excitation energy is not very sensitive to the details of the orbital order, pointing towards a dominant contribution of the crystal field. This is corroborated by the good agreement between the experimental data and the theoretical results\cite{deRay07,Solovyev06,Solovyev08,Otsuka06} for local crystal-field excitation energies and their temperature-induced relative shifts (see Tab.~\ref{Tab1}). Similar shifts of up to 40\,meV as a function of temperature are observed for the spin-flip intra-$t_{2g}$ excitation at 1.25\,-\,1.28\,eV and for the $t_{2g}^2 \rightarrow t_{2g}^1 e_g^1$ excitations at 1.84\,-\,1.88\,eV, see Fig.~\ref{Fig2} and Tab.~\ref{Tab1}.

Not only the peak energy but also the spectral weight of the low-energy intra-$t_{2g}$ excitations depends on temperature, being about 20\,-\,30\% smaller at 300\,K compared to 100\,K for $E_{\rm in}$=512.6\,eV (see Figs.~\ref{Fig4} and \ref{Fig5}) and $E_{\rm in}$=517.1\,eV (see Fig.~\ref{Fig2}). This is surprising given that the spectral weight rather increases with temperature for all higher-lying excitations, as shown in Fig.\ \ref{Fig2}. This particular behavior of the low-energy orbital excitations may be hard to understand in a pure crystal-field scenario.\cite{Ulrich2009}

\subsubsection{Dependence on scattering angle $2\theta$ or momentum $q$}

The most obvious way to distinguish a local crystal-field excitation from a propagating orbiton is to study the dependence of the orbital excitation energies on the momentum $q$. Theoretical calculations based on an effective spin-orbital superexchange model predict a significant dispersion (see Eq.~(\ref{dispersion})) with a bandwidth of up to 0.05\,eV for orbital waves propagating along the $c$ axis of YVO$_3$ in the intermediate phase with $G$-type spin order and $C$-type orbital order,\cite{Ishihara2004,Oles2007} i.e., for ferromagnetic alignment of spins along $c$. Perpendicular to $c$, the orbiton dispersion is strongly suppressed by the antiferromagnetic correlations which yield a strong renormalization of the orbiton dispersion in dimensions higher than one.\cite{Wohlfeld11} Moreover, a superexchange scenario predicts no dispersion for $q$ within the $ab$ plane if the $xy$ orbital is occupied on every site,\cite{Oles2007} which is typically assumed for YVO$_3$. This feature is due to the directional hopping properties of $t_{2g}$ orbitals, i.e., hopping between $xz$ orbitals ($yz$ orbitals) is finite only along the $x$ direction ($y$ direction) within the $ab$ plane. This completely suppresses orbital quantum fluctuations in the plane.\cite{Oles2007} The observed dispersion along the $a$ direction thus remains puzzling.

Experimentally, the $q$ dependence can be studied by varying the scattering angle $2\theta$. For small energy losses, the transferred momentum $q$\,=\,$2k_{in}\sin{\theta}$ is fully determined by the momentum of the incoming photon $k_{in}$\,=\,$2 \pi/\lambda_{in}$ and by the scattering angle $2\theta$. We have used $2\theta$\,=\,90$^{\circ}$ and $130^{\circ}$, which at $E_{\rm in}$\,=\,512.6\,eV corresponds to transferred momenta of $q_{90^{\circ}}$\,=\,0.3674\,\AA$^{-1}$ and $q_{130^{\circ}}$\,=\,0.4709\,\AA$^{-1}$, respectively. In the specular geometry of our measurement (see sketch in Fig.~\ref{Fig1}), the transferred momentum points along the $a$ axis. With a lattice constant of $a$\,=\,5.273\,\AA,\cite{Blake2002} we find $q_{90^{\circ}}/(\pi/a)$\,$\approx$\,62\,\% and $q_{130^{\circ}}/(\pi/a)$\,$\approx$\,79\,\%.

\begin{figure}[tb]
\includegraphics[width=0.99\columnwidth]{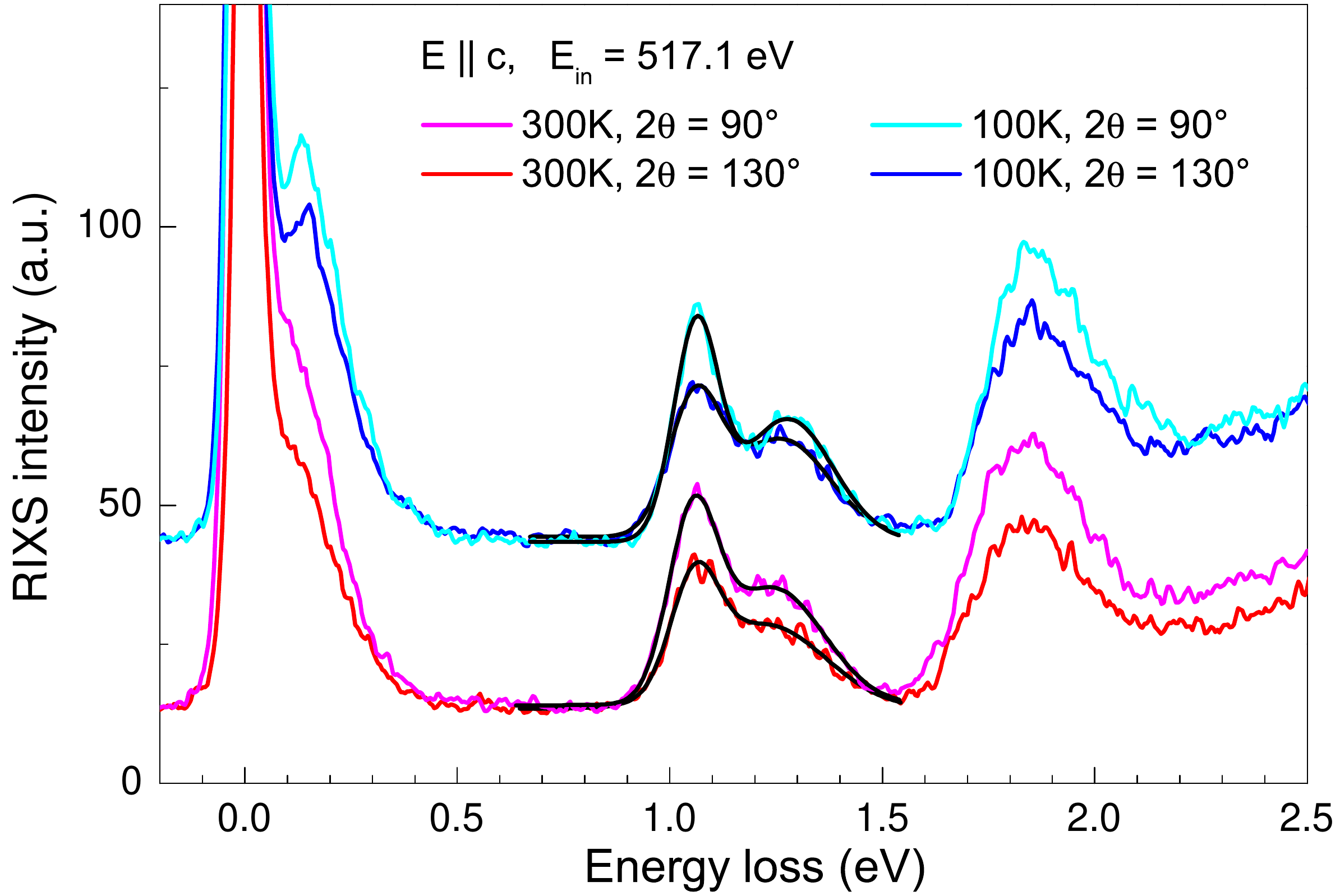}
\caption{RIXS spectra of YVO$_3$ for an incident energy of $E_{\text{in}}$\,=\,517.1\,eV at $T$\,=\,100\,K and 300\,K for $2\theta$\,=\,90$^{\circ}$ and 130$^{\circ}$. Black solid lines depict fits of the spin-flip intra-$t_{2g}$ excitations using two oscillators with Gaussian line shape.}
\label{Fig6}
\end{figure}

From a phenomenological point of view, we may compare the first moment of the RIXS feature for $2\theta$\,=\,90$^{\circ}$ and 130$^{\circ}$. At $T$\,=\,100\,K, we find a shift of 13\,meV with an estimated uncertainty of 5\,meV for $E_{\rm in}$\,=\,512.6\,eV and 517.1\,eV (see middle panel of Fig.~\ref{Fig4}). Remarkably, the same analysis in the orbitally disordered phase at 300\,K yields a shift of 6\,meV for $E_{\rm in}$\,=\,512.6\,eV but -4\,meV for 517.1\,eV, i.e., at 300\,K the first moment is independent of the scattering angle within the error bars. Fitting two Gaussians for $E_{\rm in}$\,=\,512.6\,eV yields very similar results, i.e., no shift at 300\,K (see Fig.\ \ref{Fig5} c $\&$ e) and shifts of about 20\,meV for both peaks at 100\,K (see Fig.\ \ref{Fig5} b $\&$ d). As discussed above, a bandwidth of 20\,meV appears to be plausible for YVO$_3$ with realistic parameters for the crystal field. However, a superexchange scenario predicts no dispersion for the upper peak and a finite dispersion of the lower peak only for $q \! \parallel \! c$, not for $q\! \parallel \! a$.

The RIXS feature at 0.1\,-\,0.2\,eV is composed of at least two peaks, and the shift of 13\,meV discussed above is much smaller than the line width. Therefore we have to consider an alternative scenario. A $q$ dependence of the transition matrix elements\cite{Ulrich2009} of the two different peaks may give rise to a $q$ dependence of the overall line shape (and thus of the first moment) via a transfer of the spectral weight between the two peaks around 0.1\,eV and 0.2\,eV. However, the data contain three arguments against this scenario:
(i) The first moment is independent of $2\theta$ within the error bars in the orbitally disordered phase at 300\,K.
(ii) Fits using two Gaussian peaks at 100\,K yield a very similar shift of about 0.02\,eV for both peaks.
(iii) The $2\theta$-dependence of the spectral weight of the two Gaussian peaks is opposite to the expectations of such a scenario, i.e., the spectral weight of the lower (upper) peak \textit{increases (decreases)} as the overall feature shifts to higher energy (see Fig.~\ref{Fig4}). This gives clear evidence that this alternative scenario of a transfer of spectral weight can be ruled out and supports the interpretation that the observed dependence on $2\theta$ indeed reflects an intrinsic $q$ dependence,
i.e., a finite dispersion of the low-energy orbital excitations for $q\! \parallel \! a$.

For comparison, we also determined the $2\theta$-dependence of the higher-lying RIXS peaks at 1.1-1.3\,eV by fitting two Gaussian peaks to the data for $E_{\rm in}$\,=\,517.1\,eV, see Fig.~\ref{Fig6}. Note that these peaks have hardly any RIXS intensity for $E_{\rm in}$\,=\,512.6\,eV, see Fig.~\ref{Fig1}. Thus far, a possible dispersion of these high-energy excitations has not been considered for the vanadates. The peak position of the well-pronounced lower peak at 1.07\,eV with a width of about 0.13\,eV does not depend on the scattering angle $2\theta$ within the error bars. For the less intense, broader peak at 1.28\,eV, we find that the peak position is about 15\,meV higher for $2\theta$\,=\,90$^{\circ}$ than for 130$^{\circ}$. However, this peak is much broader and shows less intensity, and the fits are affected by the noise level. Therefore, the error bars of the fitted peak position are larger, and we conclude that its position is independent of $2\theta$ within the error bars.

\subsubsection{Summary of low-energy RIXS feature}

Summarizing this section, we find evidence that neither a pure superexchange model nor a pure crystal-field scenario can explain our data. The peak energy of 0.1\,-\,0.2\,eV and its rather small temperature dependence clearly point towards a significant or possibly even dominant contribution of the crystal field. The large line width indicates a vibronic coupling to phonons and possibly also the coupling to magnons. The temperature dependence of the intensity and in particular
the finite dispersion strongly suggest that also superexchange interactions play an important role. We emphasize that our data show a finite dispersion for momentum $q\! \parallel \! a$, which is not expected in the superexchange models considered thus far.\cite{Ishihara2004,Oles2007} In order to quantify the individual contributions of the crystal field and of superexchange interactions both have to be considered on the same footing, which is a challenging task. The excitation spectrum is then dominated by an orbiton-phonon continuum.\cite{Schmidt07} If we consider also the coupling to magnons,\cite{Wohlfeld11,Wohlfeld12} we have to expect an orbiton-magnon-phonon continuum. A finite dispersion of the orbital excitations is expected to reveal itself via a momentum-dependent \textit{line shape} of this continuum.\cite{Schmidt07}
A detailed study of the $q$ dependence of the line shape remains a challenge to both theory and experiment.

\subsection{Two-orbiton peak at the O $K$ edge}\label{Kedge}

\begin{figure}[tb]
\center\includegraphics[width=0.9\columnwidth]{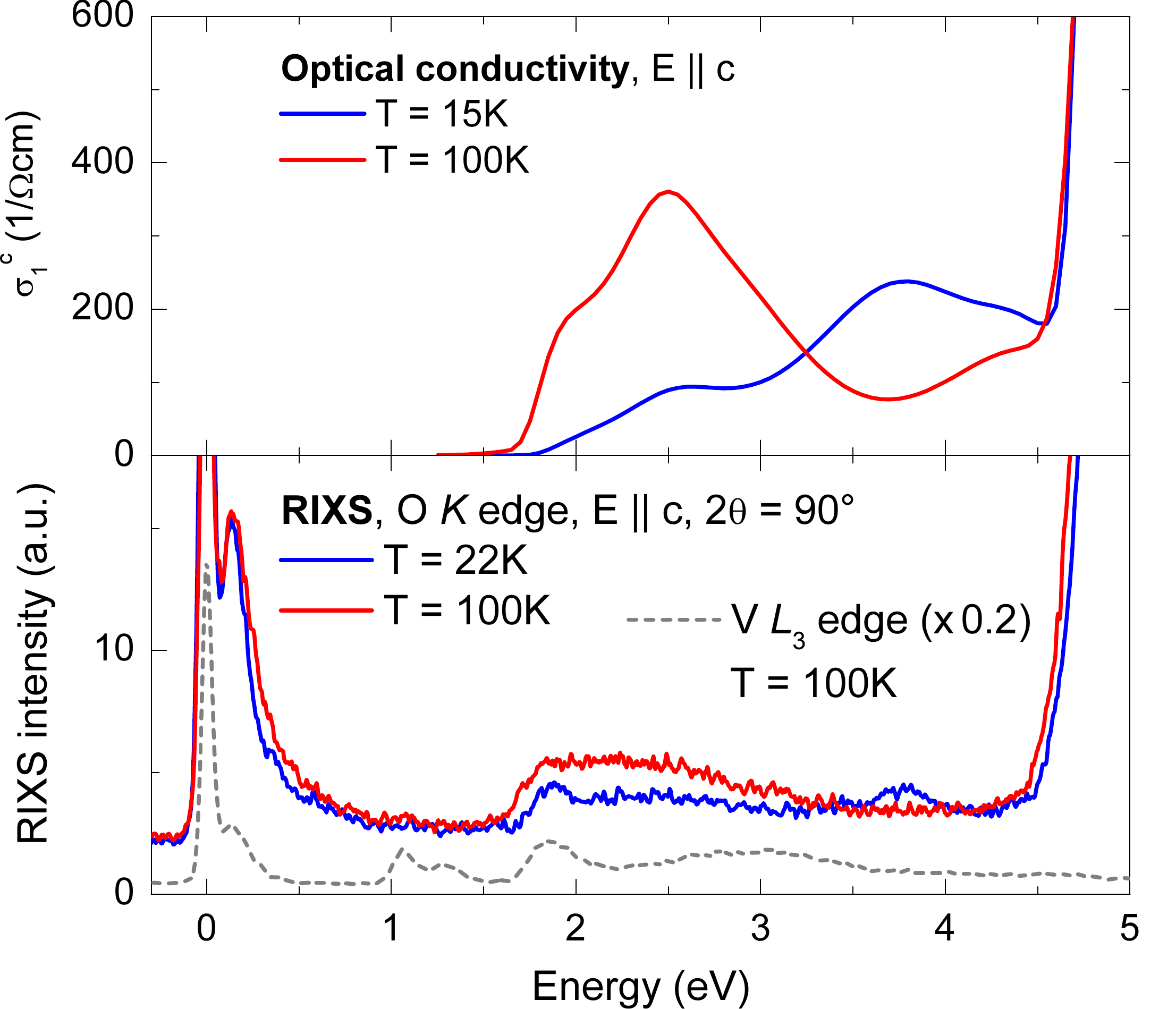}
\caption{Comparison of RIXS data measured at the O $K$ edge (bottom panel) and of the optical conductivity $\sigma_1^c(\omega)$ of YVO$_3$ as determined by ellipsometry (top panel, from Ref.~\onlinecite{Reul12}). The dashed line in the bottom panel shows data for the V $L_3$ edge from Fig.~\ref{Fig2}.}
\label{Fig7}
\end{figure}

We propose that a more direct view on the relevance of intersite or superexchange interactions at present can be obtained from the RIXS data measured at the O $K$ edge, which differs in many respects from the $L_3$ edge data, see Figs.~\ref{Fig1} and \ref{Fig7}. The spin-flip excitations at 1.1\,-\,1.3\,eV are absent (or very weak) due to the spin-selection rule, as discussed above. The low-energy RIXS feature with a peak at 0.1\,-0.2\,eV extends to much higher energies, see bottom panels of Figs.~\ref{Fig4} and \ref{Fig7}. Finally, also the peak positions above 1.5\,eV are different. The very large intensity above 4.5\,eV suggests x-ray fluorescence emission. We speculate that this strong feature above 4.5\,eV corresponds to both fluorescence and charge-transfer excitations. This is based on the astounding overall agreement between the optical conductivity\cite{Reul12} and the RIXS data from the O $K$ edge (see Fig.~\ref{Fig7}):
(i) the absorption edges of Mott-Hubbard excitations and of charge-transfer excitations are observed at about 1.7\,eV and 4.5\,eV, respectively,
(ii) the dominant contribution is a broad feature peaking at about 2.5\,eV,
(iii) between 1.7\,eV and 3.3\,eV the spectral weight is smaller in the low-temperature phase,
(iv) the peak at about 3.8\,eV is present only in the low-temperature phase,
and (v) both onset energies at about 1.7\,eV and 4.5\,eV show a similar temperature dependence.
This gives clear evidence that RIXS at the O $K$ edge is sensitive to Mott-Hubbard and charge-transfer excitations, i.e., to intersite excitations.

Additionally, we expect to observe the \textit{spin-conserving} orbital excitations discussed above for the V $L_3$ edge. In the optical conductivity, the spectral weight of orbital excitations is orders of magnitude smaller compared to the Mott-Hubbard and charge-transfer excitations shown in the top panel of Fig.~\ref{Fig7}. This explains that the pre-peak at about 1.9\,eV is more pronounced in the RIXS data than in the optical conductivity, in particular in the low-temperature phase. RIXS at the O $K$ edge shows both the Mott-Hubbard excitations and the orbital excitation $t_{2g}^2 \rightarrow t_{2g}^1 e_g^1$ at about 1.8\,-\,1.9\,eV (see Sec.~\ref{assign}). We emphasize that the RIXS intensity between 2.3\,eV and 2.7\,eV at the V $L_3$ edge was assigned to an $^1\!A_1$ ($S$\,=\,0) excited state in Sec.~\ref{assign}. This excitation involves a spin flip and thus does not contribute to RIXS at the O $K$ edge, as shown above for the comparable excitations peaking at 1.1\,-\,1.3\,eV. This strongly corroborates that the broad peak at 2.5\,eV at the O $K$ edge corresponds to Mott-Hubbard excitations and not to orbital excitations.

\begin{figure}[tb]
\center\includegraphics[width=1.0\columnwidth]{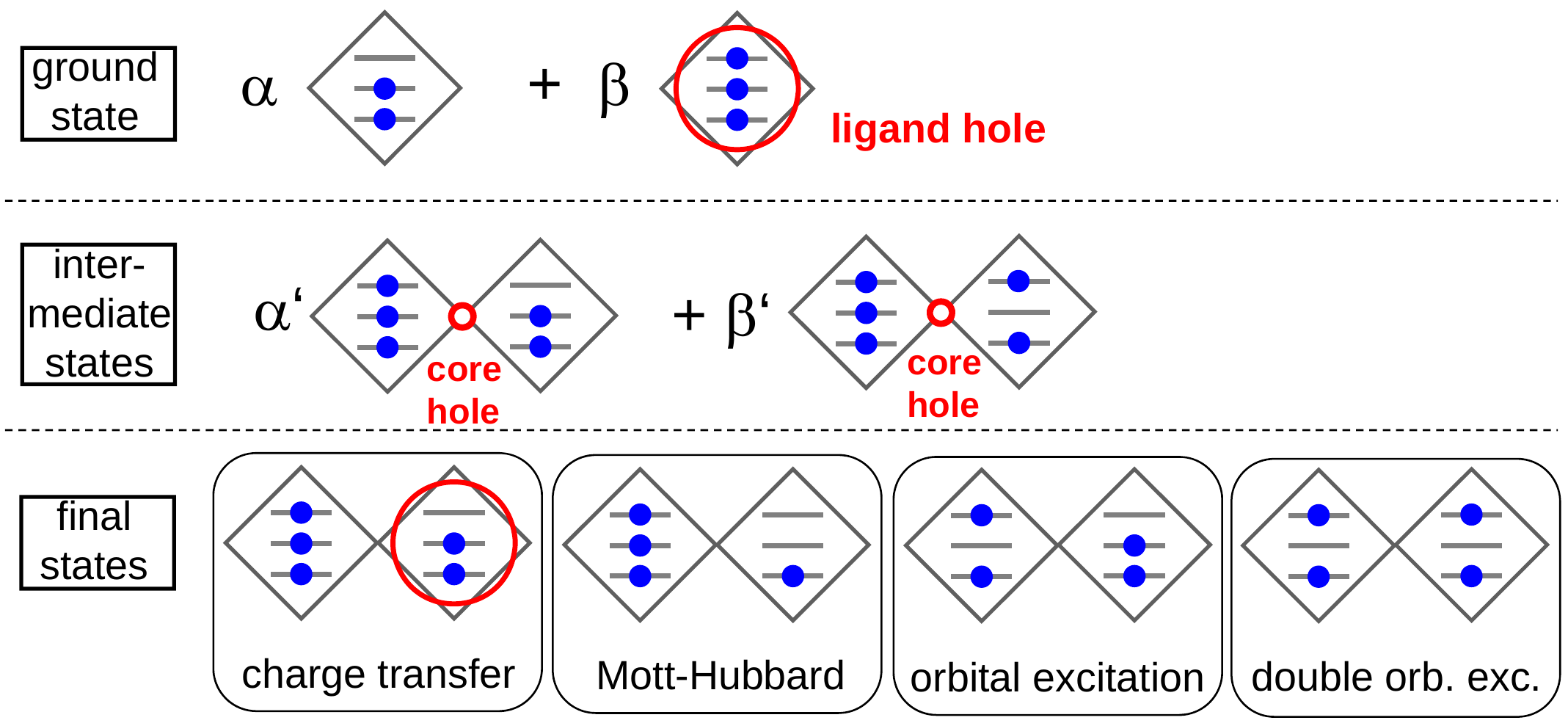}
\caption{Sketch of the ground state, intermediate states, and different final states for RIXS at the O $K$ edge of YVO$_3$. Each diamond corresponds to a VO$_6$ octahedron with three non-degenerate $t_{2g}$ levels indicated by the horizontal bars. For a single octahedron, the ground state is a superposition of $3d^2$ and $3d^3$\underline{L}, where \underline{L} denotes a ligand O $2p$ hole. See main text for a discussion of intermediate and final states.}
\label{Fig8}
\end{figure}

\begin{figure}[tb]
\includegraphics[width=0.99\columnwidth]{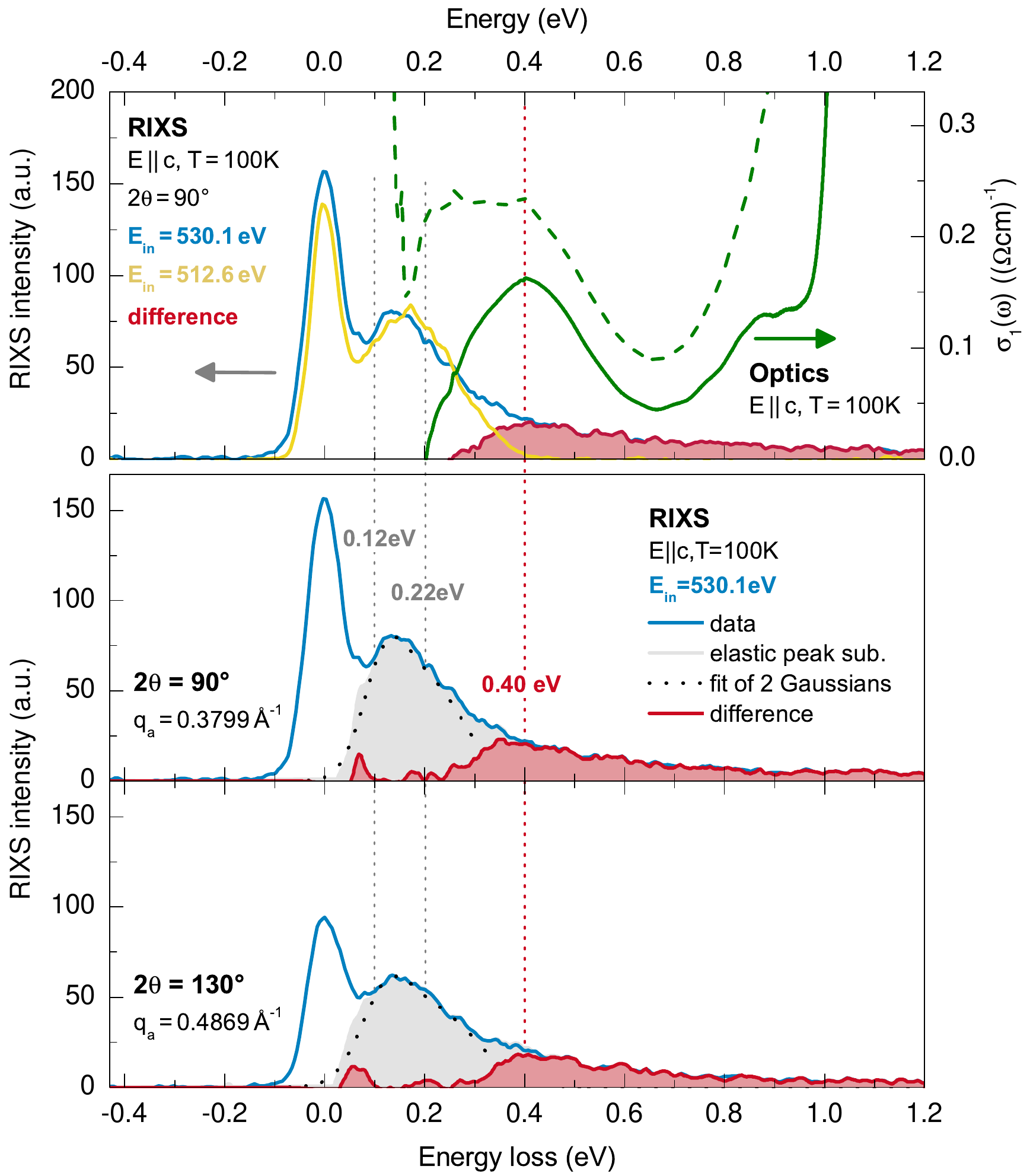}
\caption{Top panel/left axis: RIXS spectra of YVO$_3$ at $T$\,=\,100\,K and $2\theta$\,=\,90$^{\circ}$ for $E_{\text{in}}$\,=\,530.1\,eV (O $K$ edge, blue) and 512.6\,eV (V $L_3$ edge, yellow) as well as the difference between the two spectra (red). Top panel/right axis: optical conductivity $\sigma_1^c$ of YVO$_3$ at $T$\,=\,100\,K for $E$\,$\parallel$\,$c$ (dashed green) and $\Delta\sigma$\,=\,$\sigma_1^c$(100\,K)-$\sigma_1^c$(4\,K) (solid green line). The strong increase of $\sigma_1^c$ at low and high energies reflects absorption by (multi-) phonons and the onset of interband excitations, respectively (for details see Ref.\ \onlinecite{Benckiser2008}). The two-orbiton excitation is forbidden in optics below T\,=\,77\,K. Thus the 4\,K data can be used as an estimate for the background of phonons and single orbital excitations. The difference $\sigma_1^c$(100\,K)-$\sigma_1^c$(4\,K) shows an estimate of the two-orbiton contribution at 100\,K with a peak at 0.4\,eV, similar to the RIXS difference spectrum. Middle and bottom panel: RIXS spectra for $E_{\text{in}}$\,=\,530.1\,eV and two different $q$ values obtained by changing the scattering angle, i.e., 2$\theta$\,=\,90$^{\circ}$ (middle panel) and 130$^{\circ}$ (bottom panel). For an estimate of the peak shape and position of the two-orbiton peak at 0.4\,eV in RIXS, two Gaussians at 0.12 and 0.22\,eV have been fitted (width and area) after subtracting the mirrored elastic line.}
\label{Fig9}
\end{figure}

RIXS at the O $K$ edge has not been studied very much thus far. It is known to be sensitive to both orbital (or $d$-$d$) excitations and to intersite charge-transfer excitations.\cite{Ament11,Duda2000,Harada2002,Okada2006,Duda06,Guarise2010,Bisogni12,Monney2013} To the best of our knowledge, this is the first report on the observation of Mott-Hubbard excitations in RIXS at the O $K$ edge. The differences between the RIXS data from the O $K$ edge and from the V $L_3$ edge originate from the different microscopic excitation processes. In both cases, the incoming x-ray photon excites a core electron into the upper Hubbard band. At an O site, this is possible due to the hybridization between V $3d$ and O $2p$ states. In a local picture for a single VO$_6$ octahedron, the ground state can be described by $\alpha$\,|$3d^2\rangle$ +  $\beta$\,|$3d^3$\,\underline{L}$\rangle$ with $\alpha^2 + \beta^2$\,=\,1, where \underline{L} denotes a ligand O $2p$ hole (see top panel of Fig.~\ref{Fig8}). The incoming x-ray thus may promote the ligand O $2p$ hole to an O $1s$ core hole at one particular O site which connects two octahedra. This yields an intermediate state |$3d^3_i;\,1s^1;\,3d^2_j\rangle$ with a V $3d^3$ state in the upper Hubbard band, where $i$ and $j$ refer to the two V sites (see middle panel of Fig.~\ref{Fig8}). The system relaxes by filling the core hole with a valence electron, which may yield different low-energy excited states, see bottom panel of Fig.~\ref{Fig8}. The final state corresponds to a charge-transfer excitation $|3d_i^3;\, 3d_j^2$\underline{L}$\rangle$ if the core hole is filled by an O $2p$ electron from below the Fermi energy,
i.e., the final state shows both an O hole below the Fermi energy and a $3d^3$ state within the upper Hubbard band. Similarly, a Mott-Hubbard excitation results if the valence electron is taken from the lower Hubbard band, i.e., $\alpha^{\prime\prime}|3d_i^3;\, 3d_j^1\rangle + \beta^{\prime\prime}|3d_i^3;\, 3d_j^2$\underline{L}$\rangle$, which again is possible due to the $2p$-$3d$ hybridization expressed by the second term. Alternatively, an orbital excitation $|3d^{2*}_i; 3d^2_j\rangle$ is created if the core hole is filled by an electron from the upper Hubbard band, similar to the dominant mechanism at the V $L_3$ edge.

Our data were measured with linearly polarized incident x-rays with $E$\,$\parallel$\,$c$. The dipole selection rule states that the initial excitation from O $2p$ to O $1s$ involves a hole with $2p_z$ character. For this $2p_z$ hole, the $2p$-$3d$ hybridization is different for V-O-V bonds along $c$ or within $ab$. Along $c$, O $2p_z$ shows a strong hybridization with the V $e_g$ orbital with $3z^2$ symmetry. Within $ab$, O $2p_z$ hybridizes with V $t_{2g}$ orbitals with $xz$ and $yz$ symmetry. Therefore, V-O-V bonds along $c$ and within $ab$ certainly show different contributions to our O $K$ RIXS data measured with $E$\,$\parallel$\,$c$. However, a quantitative statement requires a much more thorough theoretical investigation of the microscopic processes.

Most interesting to us is the possibility to create \textit{two} orbital excitations simultaneously. In the literature, the excitation of two orbitons has been calculated for RIXS at the \textit{transition-metal} $K$ and $L$ edges with a focus on manganites and titanates.\cite{Ishihara2000,Forte08a,Ament2010} Two different mechanisms have been described. In the intermediate state, the core hole may \lq{}shake up\rq{} the valence electrons on the transition metal site, giving rise to, e.g., a single orbital flip. This state has to be translated into the eigenstates of the bulk, e.g., a local crystal-field excitation or a single (or multi-) orbiton. Alternatively, the core hole may modulate the superexchange, giving rise to two orbital flips on adjacent sites. Similarly we expect for RIXS at the O $K$ edge that the O $1s$ core hole modulates the V-O-V intersite superexchange and \lq{}shakes up\rq{} the valence electrons. Therefore, the intermediate state may evolve from $|3d_i^3$;\,$1s^1$;\,$3d_j^2\rangle$ to $|3d_i^{2*};\,1s^1;\,3d_j^3\rangle$, which may decay to a final state with two orbital excitations, $|3d_i^{2*};\,3d_j^{2*}\rangle$, see Fig.~\ref{Fig8}. We expect that the sensitivity to such double orbital excitations is enhanced at the O $K$ edge because the excitation occurs at an O site connecting two V sites, and due to hybridization the intermediate state extends over a V-O-V bond. At the V $L_3$ edge, the intermediate state is much more localized on a single V site due to the strong interaction with the V $2p$ core hole. This is supported by our observation of Mott-Hubbard excitations at the O $K$ edge at 2.5\,eV, which clearly demonstrates the sensitivity to intersite excitations. Moreover, similar intersite excitations have been reported for RIXS at the O $K$ edge of insulating late transition-metal oxides. In NiO, a double spin flip has been observed on neighboring sites,\cite{Duda06} and the excitation of two magnons has been discussed for insulating parent compounds of the high-$T_c$ cuprates.\cite{Harada2002,Guarise2010,Bisogni12} Similar to our case, the two-magnon feature in the cuprates can be revealed by comparison of RIXS at the Cu $L_3$ edge and the O $K$ edge.\cite{Guarise2010,Bisogni12}

In the optical data, a peak at 0.4\,eV for $E$\,$\parallel$\,$c$ in the intermediate monoclinic phase (see top panel in Fig.~\ref{Fig9}, right axis) has been interpreted as a two-orbiton excitation, i.e., an \textit{exchange of orbitals between two neighboring V sites} along the $c$ axis.\cite{Benckiser2008} This is equivalent to a double orbital flip, i.e., from the $xz$ orbital to the $yz$ orbital on one site and \textit{vice versa} on a neighboring site.

The RIXS spectra for $E_{\rm in}$\,=\,530.1\,eV in the bottom panels of Figs.~\ref{Fig4} and \ref{Fig7} indeed reveal an additional contribution at the O $K$ edge, which extends up to about 0.9\,eV, roughly twice the cut-off of the single orbital excitations. After subtracting an estimate of the single orbital excitations, the O $K$ edge data show a peak at about 0.4\,eV (see Fig.~\ref{Fig9}), in good agreement with the optical data. We stress that only orbital excitations can be expected in this energy range in YVO$_3$. As discussed above, phonons and magnons are relevant below 0.1\,eV only, and the Mott-Hubbard gap is larger than 1\,eV.\cite{Benckiser2008} The RIXS data at the V $L_3$ edge clearly demonstrate that the single orbital excitations are located at 0.10\,-\,0.22\,eV for all temperatures measured here, and the data from the O $K$ edge strongly corroborate the existence of the two-orbiton excitation. Together, RIXS and optics provide strong evidence for the two-orbiton interpretation of the feature at 0.4\,eV, demonstrating the relevance of orbital exchange processes in YVO$_3$.

\section{Summary and Conclusion}\label{conclusion}

In conclusion, we have reported high-resolution RIXS results at the V $L_3$ edge of YVO$_3$ which reveal the orbital excitations with unprecedented quality. By comparison with optical data,\cite{Benckiser2008} peaks at 0.1\,-\,0.2\,eV, 1.07\,eV, 1.28\,eV, 1.86\,eV, and between 2.2\,eV and 3.5\,eV unambiguously can be identified as orbital excitations. Orbital excitations to final states with either $S$\,=\,1 or $S$\,=\,0 are very well resolved, demonstrating the superior capability of high-resolution RIXS for the study of orbital excitations. The lowest orbital excitations at 0.1\,-\,0.2\,eV correspond to spin-conserving intra-$t_{2g}$ excitations. The excitation energy, its small temperature dependence, and the large line width indicate a large contribution of the crystal field. This low-energy RIXS feature can be decomposed into two different contributions at 0.1\,eV and 0.2\,eV, respectively, which is supported by the dependence of line shape and peak energy on the incident energy. Most remarkably these low-energy excitations show a shift of about 13\,-\,20\,meV as a function of the scattering angle. We argue that this reflects a finite dispersion of orbital excitations for $q\! \parallel \! a$. However, the shifts are much smaller than the line width and the superexchange models studied thus far do not predict any dispersion for $q\! \parallel \! a$. Therefore, we cannot disentangle quantitatively the contributions from superexchange and from the crystal field at this stage and call for further theoretical investigations of the low-energy orbital excitations in $t_{2g}$ systems. In particular, the microscopic mechanism for a dispersion for $q\!\parallel \! a$ needs to be understood.

RIXS at the O $K$ edge is more sensitive to intersite excitations such as charge-transfer excitations or excitations across the Mott-Hubbard gap. In particular, we find an additional contribution to the RIXS signal which peaks at about 0.4\,eV.\@ We interpret this feature as a two-orbiton excitation, i.e., an exchange of orbitals on adjacent sites, in agreement with optical results.\cite{Benckiser2008} A quantitative description of the low-energy orbital excitations in YVO$_3$ will have to consider both superexchange interactions and the coupling to the lattice.

\section*{ACKNOWLEDGEMENT}

The XAS and RIXS measurements were performed at the ADRESS beamline of the Swiss Light Source using the SAXES instrument jointly built by Paul Scherrer Institut (Switzerland), Politecnico di Milano (Italy), and \'{E}cole polytechnique f\'{e}d\'{e}rale de Lausanne (Switzerland). The authors thank N.~Hollmann, L.-H.~Tjeng, and L. Braicovich for fruitful discussions. This work was supported by the DFG via SFB 608 and partially by the U.S. Department of Energy, Office of Basic Energy Sciences, Division of Materials Science and Engineering, under Contract No.~DE-AC02-76SF00515 and by the Alexander von Humboldt Foundation.

\section{Appendix}
\subsection*{Crystal-field excitations of V$^{3+}$ ions in DySc$_{0.9}$V$_{0.1}$O$_{3}$ }

Above 1.6\,eV, orbital or $d$-$d$ excitations cannot be detected by optical spectroscopy on YVO$_3$ because they are masked by the much larger spectral weight of excitations across the Mott-Hubbard gap.\cite{Benckiser2008} As an alternative, we have chosen to study the orbital excitations of single V$^{3+}$ ions substituted into DyScO$_3$, which is isostructural to YVO$_3$. Due to the similarity of the ionic radii, the crystal field on the V site is expected to be very similar in DySc$_{0.9}$V$_{0.1}$O$_3$ and YVO$_3$. At the same time, the band gap is much larger in DyScO$_3$ than in YVO$_3$, which allows us to study the $t_{2g}^2 \rightarrow t_{2g}^1 e_g^1$ excitations. Moreover, Sc$^{3+}$ has a $d^0$ electron configuration, thus all orbital excitations of DySc$_{0.9}$V$_{0.1}$O$_3$ can be attributed to Dy $f$-$f$ or V $d$-$d$ excitations. The Dy $f$-$f$ excitations are well studied,\cite{HendersonImbusch} the peaks observed in our DyScO$_3$ data are indeed typical.\cite{BenckDiss} In particular, one does not expect any Dy $f$-$f$ excitations between about 1.7 and 2.6\,eV. The V $d$-$d$ excitations can be identified by comparing DySc$_{0.9}$V$_{0.1}$O$_3$ with DyScO$_3$.

Single crystals have been grown by the traveling-solvent floating-zone method. The purity and single-phase structure of the crystals was confirmed by powder and single-crystal x-ray diffraction at room temperature. Both compounds adopt the space group $Pbnm$ and have similar lattice constants and V-O bond lengths as RVO$_3$.
Typical dimensions of the as-grown rods are a few millimeter along all three crystallographic axes. The sample of DyScO$_3$ is transparent yellow, whereas the sample of DySc$_{0.9}$V$_{0.1}$O$_3$ is opaque and dark-green. For the transmittance measurements, we cut samples of DyScO$_3$ with [100] orientation and of DySc$_{0.9}$V$_{0.1}$O$_3$ with [001] orientation (both with the $b$ axis in-plane). These were polished to a thickness $d$ of 730\,$\pm$\,10\,$\mu$m in case of DyScO$_3$ and 1312\,$\pm$\,10\,$\mu$m in case of DySc$_{0.9}$V$_{0.1}$O$_3$. Transmittance measurements were performed between 0.1 and 3.0\,eV with linearly polarized light using a Fourier spectrometer.

\begin{figure}[b]
\center\includegraphics[width=0.9\columnwidth]{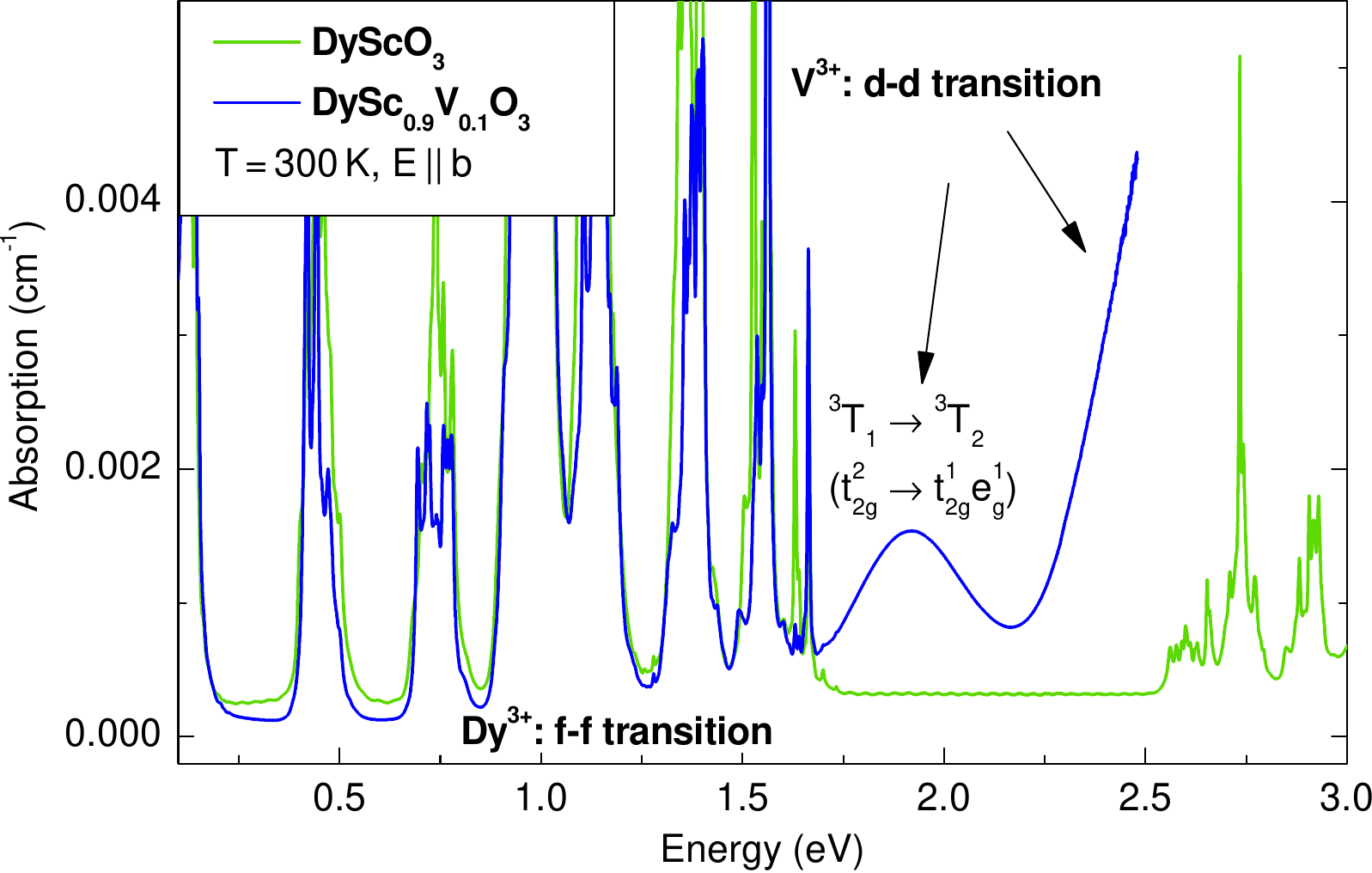}
\caption{Optical absorption of DyScO$_3$ and DySc$_{0.9}$V$_{0.1}$O$_3$ for $E$\,$\parallel$\,$b$ at 300\,K. The additional absorption peaks observed in DySc$_{0.9}$V$_{0.1}$O$_3$ are due to orbital excitations of the substituted V$^{3+}$ ions.}
\label{Fig10}
\end{figure}

In Fig.~\ref{Fig10} we show the optical absorption $\alpha = - (\ln T)/d$, where $T$ denotes the measured transmittance. At 300\,K, the $t_{2g}^2 \rightarrow t_{2g}^1 e_g^1$ absorption band of V$^{3+}$ in DySc$_{0.9}$V$_{0.1}$O$_3$ (excitation to $^3T_2$ in cubic notation) is observed at 1.92\,eV.\@ Subtracting a phonon shift of 50\,-\,80\,meV yields 1.84\,-\,1.87\,eV for the $t_{2g}^2 \rightarrow t_{2g}^1 e_g^1$ excitations, in excellent agreement with the RIXS data peaking at 1.84\,eV (see Fig.~\ref{Fig2} and Tab.~\ref{Tab1}). Observation of the lower-lying V$^{3+}$ intra-$t_{2g}$ transitions is impossible because of the superposition with the predominant $f$-$f$ transitions. Finally, the strong increase of absorption of DySc$_{0.9}$V$_{0.1}$O$_3$ above 2.2\,eV (see Fig.~\ref{Fig10}) is attributed to higher-lying  excitations, again in excellent agreement with RIXS.

%

\end{document}